\documentclass[a4paper,11pt]{article}
\usepackage{jcappub} 
\usepackage{lineno}
\usepackage{subcaption}
\usepackage{enumitem}
\nolinenumbers

\title{Multiple-cavities interferometric analysis for dark matter axions directional-sensitive search based on signal cross-correlation processing}

\author{José Reina-Valero,$^{a,*}$ Alejandro Díaz-Morcillo,$^{b}$ Benito Gimeno,$^{a}$ Antonio José Lozano-Guerrero,$^{b}$ Iván Martí-Vidal,$^{c,d}$ Juan Monzó-Cabrera,$^{b}$ Jose\:R. Navarro-Madrid,$^{b}$ Carlos Peña-Garay\,$^{e}$}
\affiliation{$^a$Instituto de Física Corpuscular (IFIC), CSIC-University of Valencia,\\
Calle Catedr\'atico Jos\'e Beltr\'an Martínez, 2, 46980 Paterna (Valencia), Spain}
\affiliation{$^b$Departamento de Tecnolog\'ias de la Informaci\'on y las Comunicaciones,\\
Universidad Polit\'ecnica de Cartagena,\\
Plaza del Hospital 1, 30302 Cartagena, Spain}
\affiliation{$^c$Departament d’Astronomia i Astrof\'isica, Universitat de Val\`encia,\\
C. Dr. Moliner 50, 46100 Burjassot, Val\`encia, Spain}
\affiliation{$^d$Observatori Astron\`omic, Universitat de Val\`encia,\\
C. Catedr\'atico Jos\'e Beltr\'an 2, 46980 Paterna, Val\`encia, Spain}
\affiliation{$^e$Laboratorio Subterráneo de Canfranc, 22880 Canfranc-Estación, Spain}

\emailAdd{Jose.Reina@uv.es}

\abstract{Current axion detection limits neglect the relevance of the relative velocity between the axion field and the detectors. However, this aspect can lead to a daily modulation of the detected axion signal. In this work, we calculate the cross-correlation of various signals potentially originated in multiple-cavity setups, and we analyze how the signal-to-noise ratio and directional sensitivity depend on the signal cross-correlation among multiple cavities. The signal-to-noise ratio after cross-correlation exhibits a greater rate of increase over time compared to the power-summation technique, making it clear that this method could be potentially employed in a real setup for the reduction of the exposure time. For the study of the daily modulation, three interferometric experiments have been proposed in this manuscript: (i) three rectangular cavities in different Earth locations; (ii) three rectangular cavities located in the same Earth spot but oriented towards different perpendicular directions; (iii) six rectangular cavities in the same Earth location but oriented towards different directions. In each set-up, we have simulated three different cavity lengths. Similar results have been found for the cases (i) and (ii): when the highest length upon the three proposed is considered, a phase difference between the recorded voltages of more than $2^{\circ}$ has been obtained with our numerical calculations. We observe a daily modulation in the imaginary part of the signals cross-correlation for experiment (iii), that could be potentially used for the characterization of the axion velocity distribution. To the knowledge of the authors, this is the first time that the cross-correlation technique has been applied to the directional sensitivity analysis of an array of haloscopes.}

\begin{document}
\maketitle
\flushbottom
\newpage
\section{Introduction}
\label{sec:intro}
The axion is probably the most elegant solution to the strong CP (Charge-Parity) problem in the paradigm of QCD  (Quantum ChromoDynamics). This solution was first proposed by Robert Peccei and Helen Quinn \cite{Peccei_Quinn, Kim:2008hd}, in order to explain the suspicious cancellation of the $\overline{\theta}$ angle, by proposing a new U(1) symmetry to the Standard Model, motivated by the idea that the QCD potential has a minimum at $\overline{\theta}$ $=$ $0$. When spontaneously broken, the Goldstone boson associated with this symmetry arises (as pointed out by Weinberg \cite{Weinberg} and Wilczek \cite{Wilczek}), being named as ``axion". It is a pseudoscalar, spin-zero particle, and probably the most promising coupling that can lead to a detection would be the coupling to photons. Via the inverse Primakoff effect \cite{Primakoff, Sikivie_2}, an axion could convert into photons under an external highly intense magnetic field, where the frequency of the generated radiation is directly related with the axion mass (KSVZ \cite{KSVZ_1, KSVZ_2} and DFSZ \cite{DFSZ_1, DFSZ_2} are two of the models that aim for explaning the relation between the coupling constant of the axion-photon decay and its mass). In addition, the axion has the added interest of being a good candidate for the main component of Dark Matter in the Universe. Due to this, and to the pessimistic results obtained for WIMPs (Weakly-Interactive Massive Particles) detection during last years, axion has become the most promising particle for explaining Dark Matter. In addition, the more general concept of the axion, the ALPs (Axion-Like Particles), cover a vast range of masses, allowing this for different techniques to be employed depending on the mass of interest. 

Several setups have been proposed for the detection of the axion-photon coupling \cite{LSW_1, LSW_2, LSW_3, LSW_4, LSW_5, Sikivie}. This work is focused on haloscopes, aiming for the detection of the axion-photon decay from the Milky Way halo axions. Different microwave haloscope experiments have been or are being performed around the world. Among others, there can be noted: ADMX \cite{ADMX}, pioneer in the detection of axions with resonant cavities; CAPP \cite{CAPP_2020, CAPP_2024}, which ruled out the low-mass region around 10 $\mu \mathrm{eV}$; HAYSTAC \cite{HAYSTAC}, sensitive to a frequency range $3.6-5.8$ GHz, employing a JPA (Josephson Parametric Amplifier) as a preamplifier; QUAX, having measured the region close to 9 GHz \cite{QUAX:2024fut}; ORGAN \cite{ORGAN}, aiming to the axion detection around a frequency of 25 GHz; RADES \cite{RADES_1, RADES_2}, this one being an European collaboration, which intends to detect the axion in the high MHz and low GHz regions; FLASH \cite{FLASH}, located at INFN Frascati National Laboratories, which is a proposal that seeks axions in the range $0.49-1.49$ $\mu\mathrm{eV}$, as well as other phenomena such as high-frequency gravitational waves, hidden photons, and chameleons; and CADEx \cite{CADEx}, hosted at the Canfranc Underground Laboratory (Huesca, Spain), aiming for the axion detection in the $90-110$ GHz range, which presents a major technological challenge due to the small wavelength size.

One of the axion features is the dependence of the frequency of the associated radiation in the axion-photon decay with the velocities with which axions are seen from the laboratory reference frame. This modification of the frequency is of order $\sim 10^{-3}c$ ($c$ being the speed of light in vacuum), and although it is typically neglected, it has been studied in some references \cite{Maxwell_Boltzman_distrib, Axion_Astronomy, directional_axion_detection}. When considering the effect of velocities, the axion field (assuming a scalar classical field due to the large occupation number of the axion) is affected by the spatial frequency term $\vec{k}_{DB}\cdot \vec{r}$, where $\vec{k}_{DB}$ is the de Broglie wavenumber and $\vec{r}$ is the position vector. Thus, it is obvious that, when considering cavities with high spatial dimensions, this effect will be more noticeable, depending also on the frequency assumed for the axion. Axion-field velocity in laboratory reference frame changes day by day, and therefore, it induces a tiny daily modulation of the axion signal, whose effect is extraordinarily small when talking of a single-cavity system. An interesting aspect to be studied is the impact of these effects when a setup with a given number of cavities is supposed. Even though the total signal of a multiple-cavity system is typically obtained by adding the individual signals (in phase) of the cavities involved, along this work we are going to introduce the effects of cross-correlation technique for the study of the axion detection and its modulation. This technique is well-known in radioastronomy \cite{synthesis_radiastronomy, tools_radiastronomy}, a field where the issue of treating with extraordinarily low SNR (signal-to-noise ratio) is very usual. Cross-correlation allows to extinguish noise in a very quick and simple way. Since in this work we will treat signals coming from different devices, cross-correlation can be useful not only for the improvement of the SNR, but also for extracting valuable information from the signal variation along the year due to changes in relative velocities between dark matter and detectors. 

In order to simulate properly the detected signals registered in the resonant cavity, the BI-RME3D (Boundary Integral - Resonant Mode Expansion 3D) method has been used.  
This technique was developed during the eighties and nineties of the last century at the Università degli Studio di Pavia (Italy), and it can be applied for obtaining the modal electromagnetic field distributions inside an arbitrarily-shaped cavity with a given number of ports connected to it, and also for obtaining the multimode scattering matrix of the resonator, reporting information about magnitude and phase of the involved output signals \cite{advanced_modal_analysis, bi-rme, bi-rme_Pablo}. This method, along with the cross-correlation technique, allows us to perform an accurate interferometric analysis of the detected signals from a multiple-cavity system, where the daily variation of the velocities of the laboratory reference frame plays a leading role.

This paper is organized as follows: first, a theoretical introduction to the BI-RME3D method is performed, explaining its fundamental features and how it can be employed for the numerical analysis of a microwave resonant cavity. Once introduced, it is explained how to simulate the resonant noise of the cavity via BI-RME3D formulation. After this, a description of the velocities involved in the movement of the laboratory reference frame is provided. Then, the detected voltage is studied, showing that the result varies when considering the de Broglie wavenumber in the calculations of the axion form factor. This leads to the proposal of the three experiments that are studied along this work with the intention of observing the directional sensitivity mentioned earlier. In addition, a study of the SNR improvement via cross-correlation technique is made, observing how the SNR evolves with the number of time averages considered in the cross-correlation and the number of cavities involved. Finally, numerical results related to the previous sections are shown and discussed (SNR improvement, phase shifts between cavities for the three proposed experiments, and the daily modulation of the cross-correlated signals). Lastly, conclusions about the study performed are provided.

\section{Theoretical formulation}

\subsection{The BI-RME3D method}

Since the mathematical formulation of this method has been previously described in several publications \cite{advanced_modal_analysis, bi-rme, bi-rme_Pablo}, here we will only outline its main features and performances. Anyway, it is important to remark at this point that this theory is directly derived from the classical Maxwell equations in frequency domain, representing one of the most relevant methods for the rigorous and accurate analysis of microwave and millimeter-wave cavities.

Considering the formulation for a cavity coupled to a single port and excited by the axion-photon decay, we can arrive to the following result:
\begin{equation}
    I_{a} = \frac{1}{\mu_{0}}g_{a\gamma\gamma}a_{0}jk\sum_{m=1}^{M}\frac{\kappa_{m}}{\kappa_{m}^2 - k^2}\left(\int_{S(1)}\vec{H}_{m}\cdot \vec{h}_{\mathrm{TEM}}\left(\vec{r}\right)\ dS\right)\left(\int_{V}\vec{E}_{m}\cdot \vec{B}_{e}\ e^{j\left(-\vec{k}_{DB}\cdot \vec{r} + \varphi\right)}\ dV\right),
    \label{eq:I_a_direccionalidad}
\end{equation}
where $I_{a}$ is the current source generated by the axion-photon coupling; $\mu_{0}$ is the vacuum magnetic permittivity; $g_{a\gamma\gamma}$ and $a_{0}$ are the axion-photon coupling constant and the axion field amplitude, respectively; $j=\sqrt{-1}$ is the imaginary unit; $k$ $=$ $\omega/c$ is the free-space wavenumber, $\omega$ being the angular frequency $\left(\omega = 2 \pi \nu\right)$; $M$ is the number of cavity resonant modes considered; $\kappa_{m}$ is the perturbed $m$-th resonant mode wavenumber (taking  into account ohmic losses by perturbing the original resonant mode wavenumbers $k_{m}$); and $\varphi$ is the intrinsic axion phase. The term $e^{j \omega t }$ has been assumed and omitted in the phasors notation. The first surface integral represents the coupling between the normalized magnetic field of the $m$-th cavity mode, $\vec{H}_{m}$, and the normalized magnetic field of the TEM mode of the coaxial probe connected to the cavity $\vec{h}_{\mathrm{TEM}}$; and the second volumen integral represents the coupling between the resonant electric field of the $m$-th cavity mode, $\vec{E}_{m}$, and the external static magnetic field $\vec{B}_{e}$, maintaining both the de Broglie and axion intrinsic phases. Additionally, the following identity is fulfilled, based on the Kirchhoff laws:
\begin{equation}
    I_{w} = I_{a} - I_{c} = I_{a} - Y_{c}V_{c}
\end{equation}
where $I_{w}$ is the current extracted from the waveguide; $Y_{c}$ is the cavity input admittance; $V_{c}$ is the voltage in the cavity; and $I_{c}$ is the current flowing towards the cavity, as it is shown in Figure \ref{fig:circuito_equivalente_axion}. Thus, we have derived a simple circuital equation that can be understood with basic knowledge of classical network analysis: the presence of a charged current density from a certain source (in this case, the presence of the axion field) generates a current $I_{a}$ that is splitted into two parts: one part goes to generating a current that is delivered to the waveguide (connected to the detector), $I_w$, and the other one goes to the resonant cavity itself (and dissipated by Joule effect within the cavity), $I_{c}$. Applying classical network theory, we can easily note that:
\begin{equation}
    V_{c} = \frac{I_{a}}{Y_{w} + Y_{c}},
\end{equation}
where $Y_{c}$ can be expressed, as any admittance, by a real and an imaginary part, $Y_{c} = G_{c} + j B_{c}$; and $Y_{w}$ $=$ $1/Z_{w}$, where $Z_{w}$ is the impedance of the connected waveguide port. As a consequence, the power consumed by the cavity $P_{c}$ and the power delivered to the waveguide $P_{w}$ can be easily calculated as follows:

\begin{figure}[h]
    \centering
    \includegraphics[scale = 0.6]{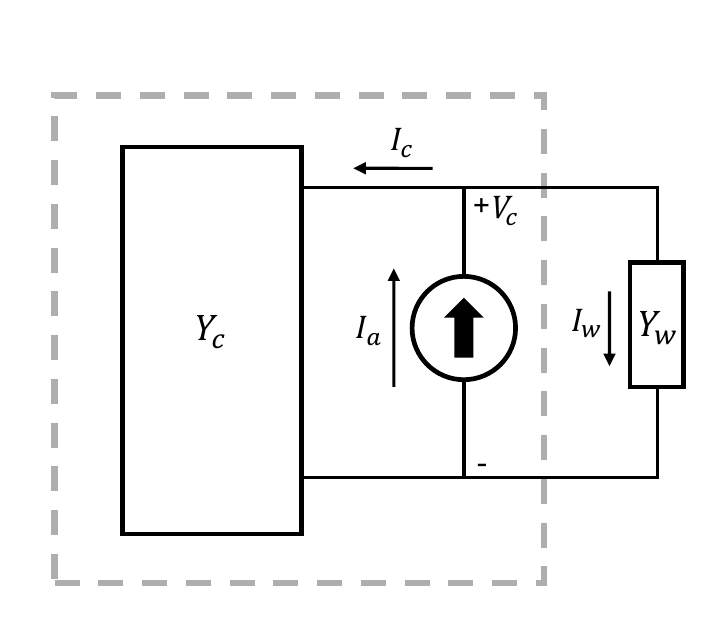}
    \caption{Network that represents, in a simple way, the implications of the axion decay inside our resonant cavity \cite{bi-rme_Pablo}. $I_{a}$ is the current generated by the axion density current, which acts as a current source. The voltage $V_{c}$ is directly associated with the one that would be measured in a conventional experiment.}
    \label{fig:circuito_equivalente_axion}
\end{figure}

\begin{equation}
    P_{c} = \frac{1}{2}Re\left(V_{c}I_{c}^{*}\right),
\end{equation}
\begin{equation}
    P_{w} = \frac{1}{2}Re\left(V_{c}I_{w}^{*}\right).
\end{equation}

The fundamental conclusion of the explained formalism is that the BI-RME3D formulation allows to establish a connection between the well-known problem of the axion decay to a photon and the classical electromagnetic network theory of a microwave resonator, which is depicted in Figure \ref{fig:circuito_equivalente_axion}. In addition, we would like to remark that $I_{a}$ and $V_{c}$ are phasors, thus having their associated amplitude and phase. This implies that BI-RME3D allows to calculate the phase of the signal that would be measured in a particular experiment, enabling us to extend the mathematical analysis of the signal beyond its amplitude.

\subsection{Resonant noise simulation by means of the BI-RME3D method}

Resonant noise can be simulated through a transference function as it is commonly done in bibliography \cite{Sikivie_2,australianos_IEEE}. However, in order to maintain the same formulation along this work, we have introduced the resonant noise in the BI-RME3D theory.

The noise that is extracted from the cavity in a measurement can be described in terms of an equivalent random electric current density $\vec{J}_{n}$. Thus, the current generated inside a cavity excited by noise follows the next expression:
\begin{equation}
I_{n} =  -\sum_{m=1}^{M}\frac{\kappa_{m}}{\kappa_{m}^2 - k^2}\left(\int_{S(1)}\vec{H}_{m}\cdot \vec{h}_{\mathrm{TEM}}\left(\vec{r}\right)\ dS\right)\left(\int_{V}\vec{E}_{m}\left(\vec{r}\ ^\prime\right)\cdot \vec{J}_{n}\ dV^\prime\right).
\label{eq:I_n}
\end{equation}

\begin{figure}[h]
    \centering
    \includegraphics[scale = 0.6]{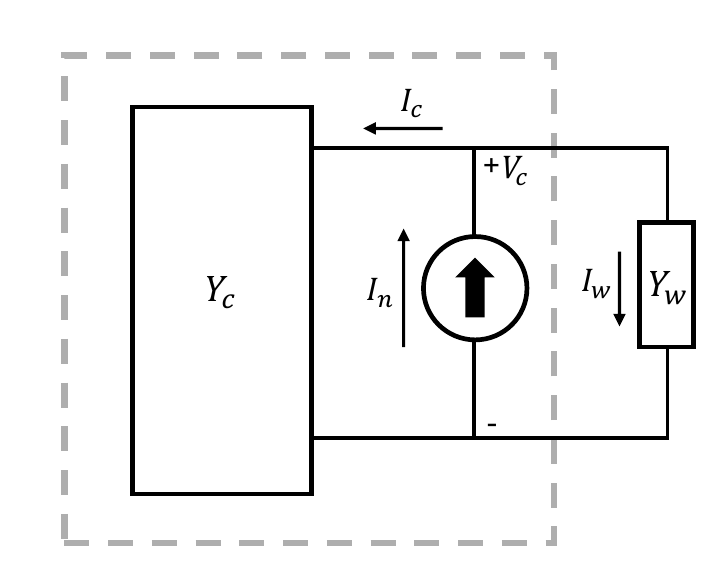}
    \caption{Equivalent circuit that shows how to add the resonant thermal noise in the BI-RME3D formalism.}
    \label{fig:circuito_equivalente_solo_ruido}
\end{figure}

As it can be shown in Figure \ref{fig:circuito_equivalente_solo_ruido}, the recorded voltage would change consequently to $V_c = I_{n}/\left(Y_{w} + Y_{c}\right)$.
Initially, the value of $\vec{J_{n}}$ is taken to follow a random Gaussian distribution. However, an extra condition must be imposed in order to obtain the proper result. Considering only the thermal noise in the system, namely an empty cavity without axions, the power delivered to the waveguide ($P_{w}$) must be properly normalized to the noise expected power $P_{n}$ according with (see Figure \ref{fig:circuito_equivalente_solo_ruido}),
\begin{equation}
    P_{w} = \frac{1}{2}Re\left(V_{c}I_{w}^{*}\right) = \frac{1}{2}\frac{|I_{n}|^2}{|Y_{c} + Y_{w}|^2}\mathrm{Re}\left(Y_{w}^*\right)\,; P_{n} = k_{B}T\Delta \nu  \rightarrow P_{n} = P_{w} \Rightarrow |I_{n}|^2 = 2P_{n}\frac{|Y_{c} + Y_{w}|^2}{\mathrm{Re}\left(Y_{w}^*\right)}.
\end{equation}
Here, $k_{B}$ represents the Boltzmann constant, $T$ represents the system temperature expressed in K, and $\Delta \nu$ is the receiver bandwidth, which coincides with the axion bandwidth in order to introduce as less noise as possible. This normalization is necessary because the expected noise power is the sole information available about the noise of the system. 

\begin{figure}[h]
    \centering
    \includegraphics[scale = 0.6]{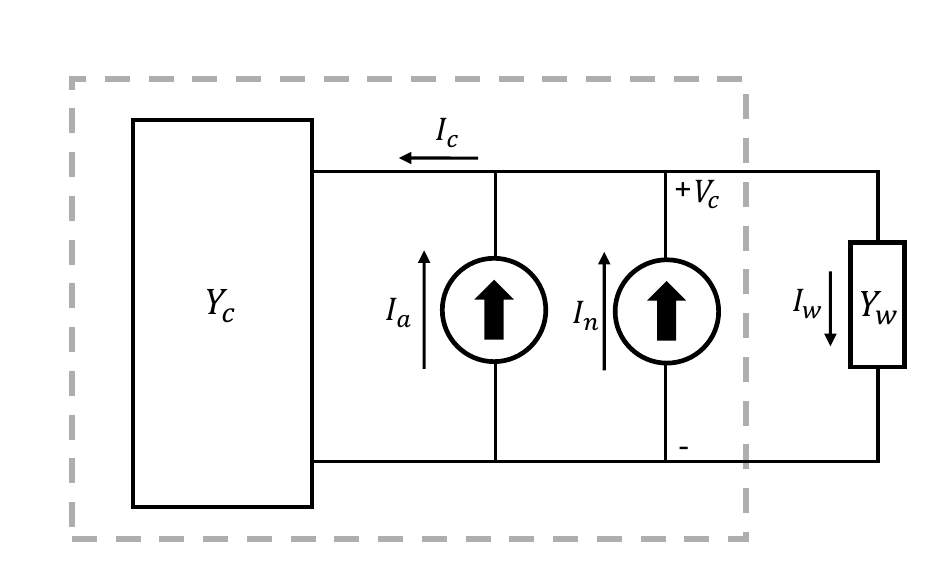}
    \caption{Equivalent circuit that shows how to add the resonant noise in the BI-RME3D formalism. Here, the axion current and the intrinsic noise of the cavity can be modeled as two current sources.}
    \label{fig:circuito_equivalente_axion_ruido}
\end{figure}

If both the axion and the resonant noise are considered, the total equivalent current $I_{T}$ adopts the following form:
\begin{equation}
    \begin{aligned}
        I_{T} & =  -\sum_{m=1}^{M}F_{m1}^{(1)}\frac{\kappa_{m}}{\kappa_{m}^2 - k^2}\int_{V}\vec{E}_{m}\left(\vec{r}\ ^\prime\right)\cdot \left(\vec{J}_{a}\left(\vec{r}\ ^\prime\right) +  \vec{J}_{n}\right)\ dV^\prime =  \\
        & = -\sum_{m=1}^{M}F_{m1}^{(1)}\frac{\kappa_{m}}{\kappa_{m}^2 - k^2}\left(\int_{V}\vec{E}_{m}\left(\vec{r}\ ^\prime\right)\cdot \vec{J}_{a}\left(\vec{r}\ ^\prime\right)\ dV^\prime +  \int_{V}\vec{E}_{m}\left(\vec{r}\ ^\prime\right)\cdot \vec{J}_{n}\ dV^\prime\right) = \\ & = I_{a} + I_{n},
        \label{eq:I_a_plus_I_n}
    \end{aligned}
\end{equation}
whose equivalent circuit would be the one shown in Figure \ref{fig:circuito_equivalente_axion_ruido}. In Eq.\ (\ref{eq:I_a_plus_I_n}), $F_{m1}^{\left(1\right)}$ is the coupling integral between the magnetic fields $\vec{H}_{m}$ and $\vec{h}_{\mathrm{TEM}}$ that appears in Eqs.\ (\ref{eq:I_a_direccionalidad}) and (\ref{eq:I_n}). 

\subsection{Description of the axion halo in laboratory coordinates}

A fundamental point is the calculation of the axion velocity in the laboratory reference frame. This velocity will be composed by two fundamental terms: one relative to the intrinsic velocity of axions in the galactic halo, and another one relative to the laboratory reference frame movement with respect to them. The first one will be simulated as it is commonly made in the bibliography, i.e., assuming a Maxwell-Boltzmann distribution \cite{Maxwell_Boltzman_distrib, HAYSTAC_analisis}. In this way, in the Milky Way reference frame, the Maxwellian distribution has the next form:
\begin{equation}
    fd^{3}v = \left[\frac{\beta}{\pi}\right]^{3/2}e^{-\beta v^2}d^{3}v,
\end{equation}
where $f$ is the probability density function; the differentials have been rewritten as $d^{3}v$ $\equiv$ $dv_{x}dv_{y}dv_{z}$, and $\beta$ $=$ $m_{a}/\left(2k_{B}T\right)$, $m_{a}$ being the axion mass. In order to transform this expression to the laboratory reference frame where our experimental setup is operating, we will need to take into account the velocities of our reference frame with respect to the Milky Way reference system. Specifically, four velocities will be involved:
\begin{equation}
\vec{v}_{\mathrm{lab}} = \vec{v}_{\mathrm{LSR}} + \vec{v}_{\mathrm{pec}} + \vec{v}_{\oplus} + \vec{v}_{\mathrm{rot}},
    \label{velocidades}
\end{equation}
where $\vec{v}_{\mathrm{LSR}}$ (Local Standard of Rest) refers to the velocity with respect to the center of our galaxy of the galactic material that surrounds the Sun up to approximately 100 pc \cite{v_LSR_1, v_LSR_2}; $\vec{v}_{\mathrm{pec}}$ is the velocity of the Sun with respect to the mentioned galactic material \cite{v_pec}; $\vec{v}_{\oplus}$ is the revolution velocity of the Earth around the Sun \cite{v_Earth}; and $\vec{v}_{\mathrm{rot}}$ is the linear rotation velocity of the Earth with respect to its own rotation axis. However, an important detail is that, although we know the shape and value of these vectors, each of them is expressed, for convenience, in a different coordinate system. Thus, $\vec{v}_{\mathrm{LSR}}$, $\vec{v}_{\mathrm{pec}}$ and $\vec{v}_{\oplus}$ are originally described in galactic coordinates and $\vec{v}_{\mathrm{rot}}$ in laboratory coordinates. In order to correctly perform the sum we must transform the velocities expressed in galactic coordinates into laboratory coordinates, and for this issue we must use the equatorial coordinates as an auxiliary reference frame. Therefore, the correct expression of the sum of these four velocities would be:
\begin{equation}
    \vec{v}_{lab} = \mathrm{R}_{lab}\mathrm{R}_{gal}\left(\vec{v}_{\odot} + \vec{v}_{\oplus}\right) + \vec{v}_{\mathrm{rot}},
\end{equation}
where $\mathrm{R}_{gal}$ is the transformation matrix from galactic to equatorial coordinates, $\mathrm{R}_{lab}$ is the transformation matrix from equatorial to laboratory coordinates, and, since both $\vec{v}_{\mathrm{LSR}}$ and $\vec{v}_{\mathrm{pec}}$ refer to the movements around the Milky Way galactic center and they are constant in galactic coordinates, it is convenient to include them in one so that $\vec{v}_{\odot}$ $=$ $\vec{v}_{\mathrm{LSR}}$ $+$ $\vec{v}_{\mathrm{pec}}$. With this consideration, the final expression of the velocity vector in laboratory coordinates is \cite{directional_axion_detection}:
\begin{equation}
    \vec{v}_{lab}\left(t\right) = \begin{pmatrix}
    -\sigma_{1}\sin\left(\lambda_{lab}\right)\cos\left(\tau_{d}\right) - \sigma_{2}\sin\left(\lambda_{lab}\right)\sin\left(\tau_{d}\right) + \sigma_{3}\cos\left(\lambda_{lab}\right)\\
    \sigma_{1}\sin\left(\tau_{d}\right) - \sigma_{2}\cos\left(\tau_{d}\right) - v_{\mathrm{rot}}\cos\left(\lambda_{lab}\right)\\
    \sigma_{1}\cos\left(\lambda_{lab}\right)\cos\left(\tau_{d}\right) + \sigma_{2}\cos\left(\lambda_{lab}\right)\sin\left(\tau_{d}\right) + \sigma_{3}\sin\left(\lambda_{lab}\right),
    \end{pmatrix}
    \label{matrizfinal}
\end{equation}
where $\lambda_{lab}$ is the latitude of the observer on Earth; $\tau_{d}$ is the sidereal time, that can be expressed as $\tau_{d}$ $=$ $\omega_{d}\left(t - t_d\right)$ $+$ $\phi_{lab}$, $t$ being expressed in days of the year starting from January 1st, $\omega_{d}$ $=$ $2\pi /\left(0.9973\, \mathrm{days}\right)$, $t_d$ $=$ $0.721$ days, and $\phi_{lab}$ is the terrestrial longitude; and $\sigma_{n}$ is the result of multiplying the row $n$ of the matrix $R_{gal}$ by the $n$ component of the sum of velocities $\vec{v}_{\odot} + \vec{v}_{\oplus}$. In this way we see that the dependence on the latitude comes not only from $\vec{v}_{\mathrm{rot}}$ as it would be expected, but also from the rotation matrix. Thus, taking into account that $\|\vec{v}_{\odot}\|$ $\sim$ $10^5$ m/s and that $\|\vec{v}_{\mathrm{rot}}\|$ $\sim$ $10^2$ m/s, the contribution to the latitude that $\vec{v}_{\odot}$ gives is three orders of magnitude higher than the one of $\vec{v}_{\mathrm{rot}}$. Thus, we can observe that the value of our vector of velocities varies with $\lambda_{lab}$, so we will consider in the simulations four different scenarios: the first one, where velocities are not accounted; in the second one, we will assume that the experimental testbed is placed in the North Pole ($\lambda_{lab}$ $=$ $90^{\circ}$); the third one, in the LSC (Laboratorio Subterráneo de Canfranc) ($\lambda_{lab}$ $=$ $42.77^{\circ}$, $\phi_{lab}$ $=$ $-0.529^{\circ}$); and the fourth one, in the Equator ($\lambda_{lab}$ $=$ $0^{\circ}$).
In addition, we see that this velocity varies sinusoidally with time as we would expect initially, so this contribution will also be taken into account since the signal that we might detect will vary depending on the stipulated day of the year. One aspect that should be pointed out is that the laboratory reference frame was chosen in a way such that Cartesian axes $\{x,y,z\}$ point out towards $\{\mathrm{North}, \mathrm{West}, \mathrm{Zenith}\}$ directions, respectively.

Considering both effects previously mentioned (intrinsic movement of the axion and laboratory reference frame movement), it can be deduced that the probability distribution of the axion frequency has the form \cite{Maxwell_Boltzman_distrib}\cite{HAYSTAC_analisis}:
\begin{equation}
    f\left(\nu\right) = \frac{2}{\sqrt{\pi}}\left(\sqrt{\frac{3}{2}}\frac{1}{r}\frac{1}{\nu_{a}\langle\beta^2\rangle}\right)\sinh{\left(3r\sqrt{\frac{2\left(\nu - \nu_{a}\right)}{\nu_{a}\langle\beta^2\rangle}}\right)}\, \mathrm{exp}{\left(-\frac{3\left(\nu - \nu_{a}\right)}{\nu_{a}\langle\beta^2\rangle} - \frac{3r^2}{2}\right)},
    \label{eq:MB_distribution}
\end{equation}
where $\langle\beta^2\rangle$ $=$ $\langle v_{\mathrm{MB}}^2\rangle/c^2$, being $\langle v_{\mathrm{MB}}^2 \rangle$ $=$ $3k_{\mathrm{B}}T/m_{a}$; and $r$ $=$ $v_{lab}/\sqrt{\langle v_{\mathrm{MB}}^2\rangle}$. This probability distribution is the one that generates the well-known quality factor of the axion $Q_{a}$ $\sim$ $10^6$; the shape of this shifted Maxwellian is represented in Figure \ref{fig:MB_distribution}. Eq.\ (\ref{eq:MB_distribution}) is deduced in Appendix \ref{app:MB_distribution}.

\begin{figure}
    \centering
    \includegraphics[scale = 0.47]{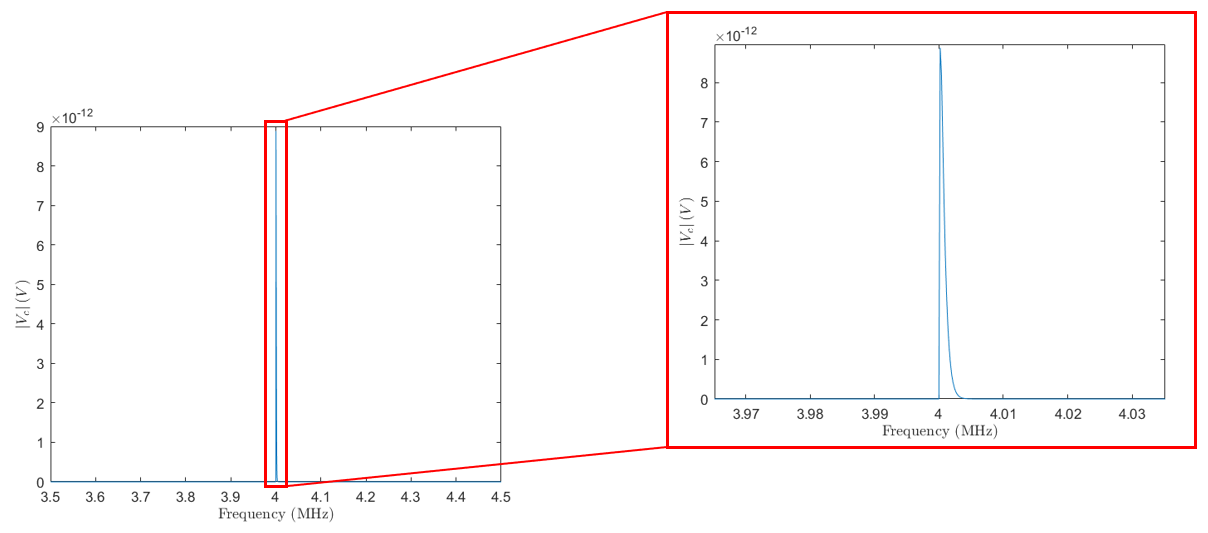}
    \caption{Simulation of the frequency distribution of a 1 GHz axion, lowered to 4 MHz via a local oscillator. Above all, its marked narrowness stands out, giving the axion a high quality factor in comparison with the ones of the cavities employed to detect it.}
    \label{fig:MB_distribution}
\end{figure}

\section{Experimental design}

The proposed technique is here studied in three rectangular cavities tuned to $\nu_{res} = 1$ GHz in which the $\mathrm{TE}_{10l}$ normalized resonant mode will be excited, where $l$ is the number of variations in the third coordinate, as it is showed in Table \ref{tab:lengths}.

\begin{figure}[h]
    \centering
    \includegraphics[scale = 0.65]{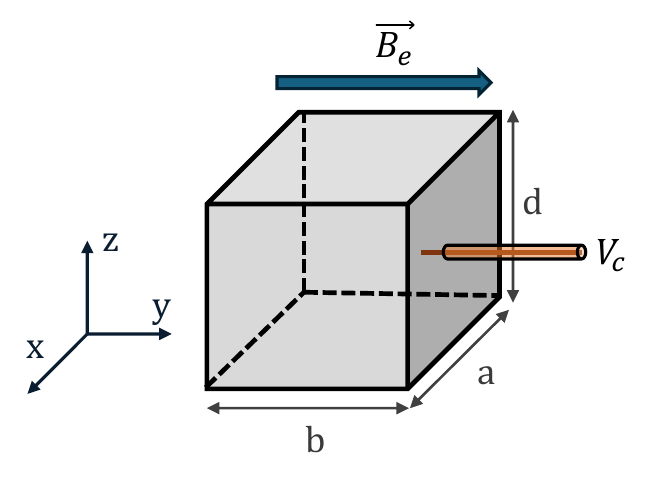}
    \caption{Single cavity setup. In orange, coaxial port and cable where the voltage $V_{c}$ is measured; in blue, external applied homogeneous magnetostatic field oriented towards $y$ axis.}
    \label{fig:setup_cavidad}
\end{figure}

\begin{table}[htbp]
\centering
\begin{tabular}{c | c | c}
\hline
Cavity & Length $d$ (mm) & $l$ value\\
\hline
C1 & 188.31 & 1\\
C2 & 2071.39 & 11\\
C3 & 3954.47 & 21\\
\hline
\end{tabular}
\caption{Lengths and $l$ values of the three rectangular cavities studied in this work.\label{tab:lengths}}
\end{table}

A pictorial representation of the single cavity setup has been made in Figure \ref{fig:setup_cavidad}. Height $a$ and width $b$ of the cavities are standardised to a WR-975 waveguide, adopting values of $a$ $=$ $247.65$ $\mathrm{mm}$ and $b$ $=$ $123.825$ $\mathrm{mm}$. Critical coupling is considered along this work (meaning that half of the energy generated by the axion within the cavity is extracted and the other half is consumed by the cavity, it is, $P_{w}$ $=$ $ P_{c}$), with an external magnetostatic field of $B_{e}$ $=$ $10$ T, and the signal will be down-converted to 4 MHz via a mixer using a conventional microwave local oscillator.

The consideration of the axion velocity in our reference frame affects the detected voltage $V_{c}$ because the de Broglie wavenumber influences the axion current $I_{a}$, as it can be demonstrated by means of the BI-RME3D method. Recovering the expression (\ref{eq:I_a_direccionalidad}), we next introduce the analytic form of the analyzed mode $\vec{E}_{10l}$. We want to remark that in order to obtain a concise result we need to truncate the summatory over modes to the studied $\mathrm{TE}_{10l}$ mode.

The normalized electric field of $\mathrm{TE}_{10l}$ follows \cite{advanced_modal_analysis}:
\begin{equation}
    \vec{E}_{10l} = -\frac{2}{\sqrt{abd}}\sin\left(\frac{\pi x}{a}\right)\sin\left(\frac{l\pi z}{d}\right)\hat{y},
\end{equation}
and after performing the integral and some algebra manipulation, a compact expression for the detected voltage can be found :
\begin{equation*}
    V_{c} = g_{a\gamma\gamma}\,\sqrt{\frac{\varepsilon_0}{\mu_0}}\,\omega\, a_0\, F_{11}\, (-1)^l \,\frac{16}{\sqrt{abd}}\,B_{e}\,\sqrt{\frac{Q_{10l}\left(Q_{10l} - 1\right) + 1/2}{\xi^2 + \zeta^2}}\cos\left(\frac{k_{DB,x}\,a}{2}\right)\cdot
\end{equation*}
\begin{equation*}
    \cdot \cos\left(\frac{k_{DB,z}\,d}{2}\right)\sin\left(\frac{k_{DB,y}\,b}{2}\right)\frac{l\pi^2\, a\,d}{\left(\pi^2 - k_{DB,x}^2\,a^2\right)\left(l^2\pi^2 - k_{DB,z}^2\,d^2\right)k_{DB,y}\sqrt{\left(Y_{w} + G_{c}\right)^2 + X_{c}^2}}\cdot
\end{equation*}
\begin{equation}
    \cdot e^{j\left[\frac{1}{2}\left(k_{DB,x}\,a\, +\, k_{DB,y}\,b\, +\, k_{DB,z}\,d\right)\, +\, \left(\varphi\, +\, \pi/2\right)\, -\, \arctan\left(\theta\right)\right]},
    \label{eq:V_c_final}
\end{equation}
where $F_{11}$ is the coupling integral between the magnetic field of the axion mode of the cavity (i.e.\ the one that we are considering) and the fundamental mode of the port; $Q_{10l}$ is the unloaded quality factor of the mode; and $\zeta$, $\xi$ and $\theta$ are parameters defined for the sake of compactness of this expression as follows,
\begin{equation}
    \zeta = k_{10l}\left(1 - \frac{1}{2Q_{10l}}\right); \ \ \ \ \ \ \ \xi = Q_{10l}\,k_{10l}\left(1 - \frac{1}{2Q_{10l}}\right)^2 - \frac{k_{10l}}{4\,Q_{10l}} - \frac{Q_{10l}}{k_{10l}}\,k^2;
\end{equation}
\begin{equation}
    \theta = \frac{X_{c}\left(\zeta + \xi\left(2Q_{10l} - 1\right)\right) - \left(Y_{w} + G_{c}\right)\left(\xi - \zeta\left(2Q_{10l} - 1\right)\right)}{X_{c}\left(\zeta - \xi\left(2Q_{10l} - 1\right)\right) + \left(Y_{w} + G_{c}\right)\left(\xi + \zeta\left(2Q_{10l} - 1\right)\right)}.
\end{equation}

\begin{figure}[t]
    \centering
    \includegraphics[scale = 0.5]{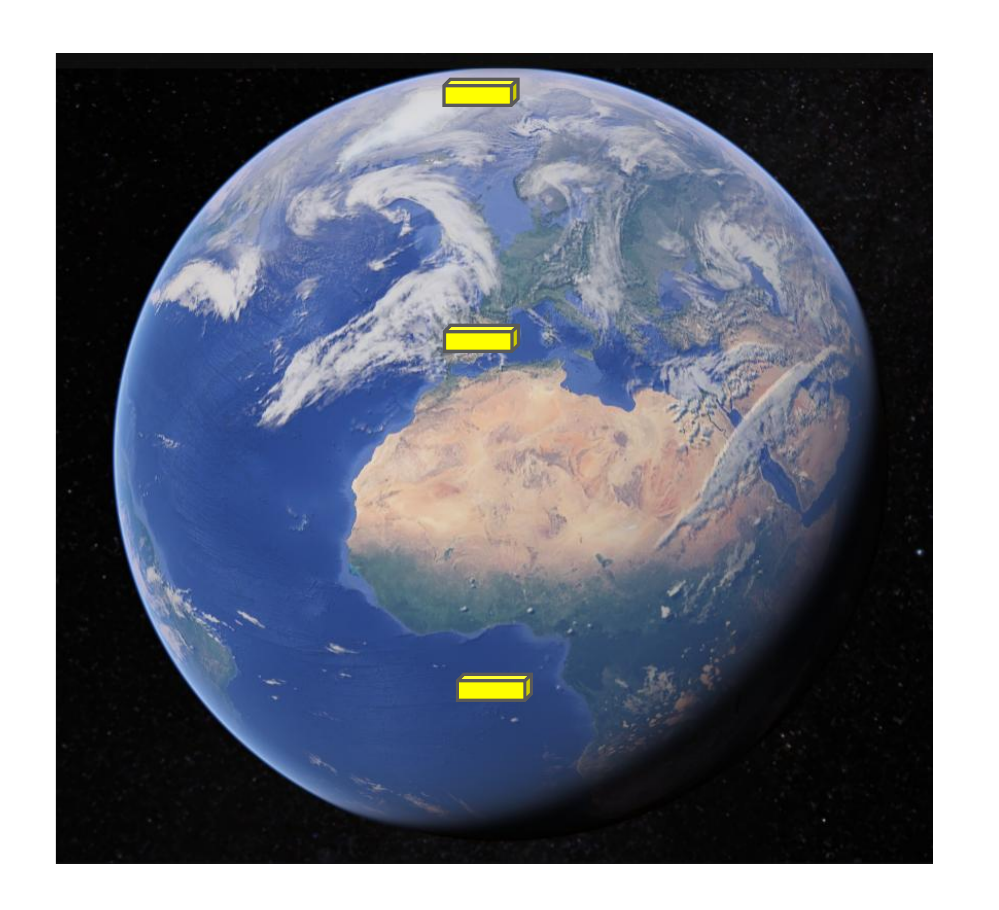}
    \caption{First experimental setup proposed in this work for the analysis of velocities in an axion detection experiment. In yellow, three rectangular cavities are placed at the same longitude ($\phi_{lab}$ $=$ $-0.38^{\circ}$) and three different latitudes: North Pole, $\lambda_{\mathrm{NP}}$ $=$ $90^{\circ}$; Canfranc Underground Laboratory, $\lambda_{\mathrm{LSC}}$ $=$ $42^{\circ} 43'$; and the Equator, $\lambda_{\mathrm{Eq}}$ $=$ $0^{\circ}$.}
    \label{fig:earth_setup}
\end{figure}

\begin{figure}[h]
    \centering
    \includegraphics[scale = 0.35]{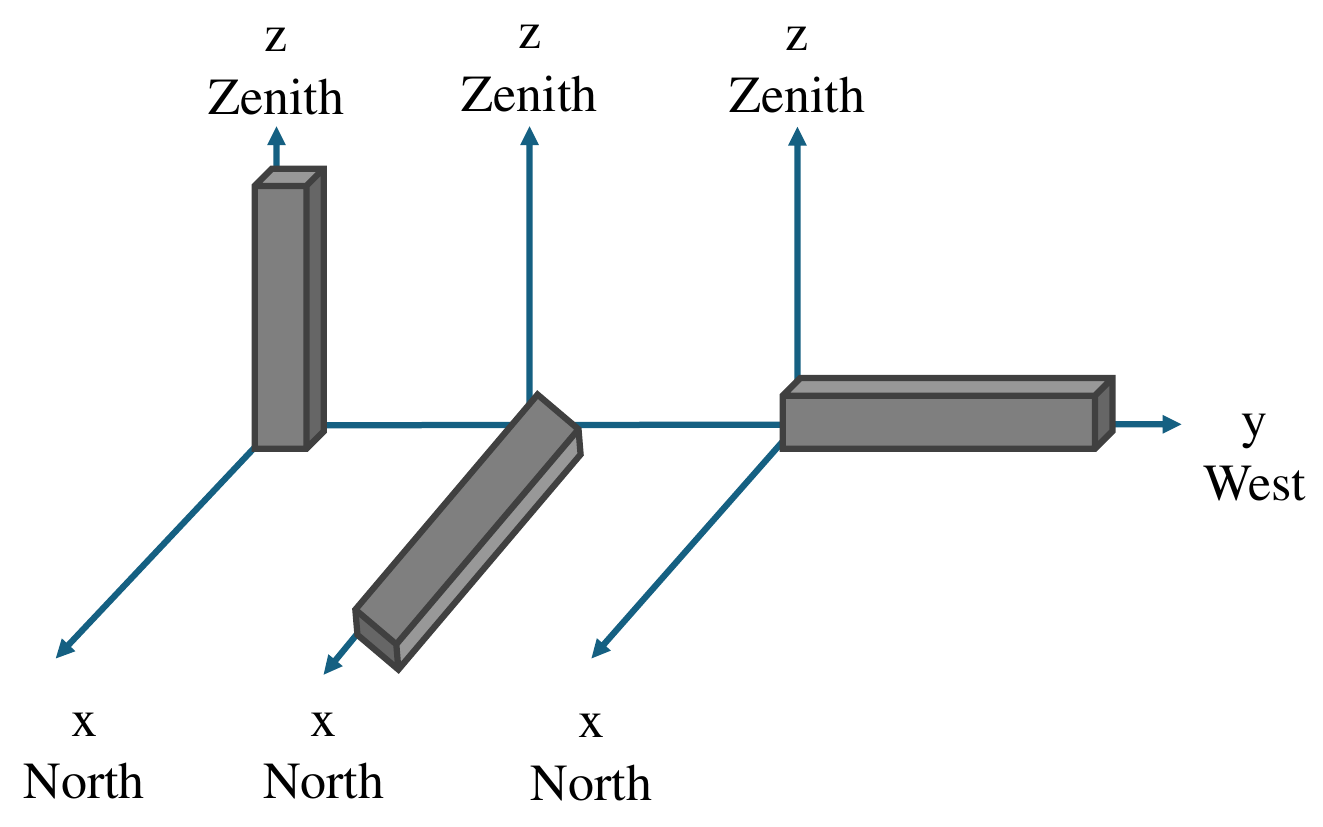}
    \caption{Second setup proposed in this work. Three rectangular cavities oriented towards the three Cartesian axes in the laboratory reference frame thus positioned in the same location.}
    \label{fig:three_oriented_cavities}
\end{figure}

\begin{figure}
    \centering
    \includegraphics[scale = 0.35]{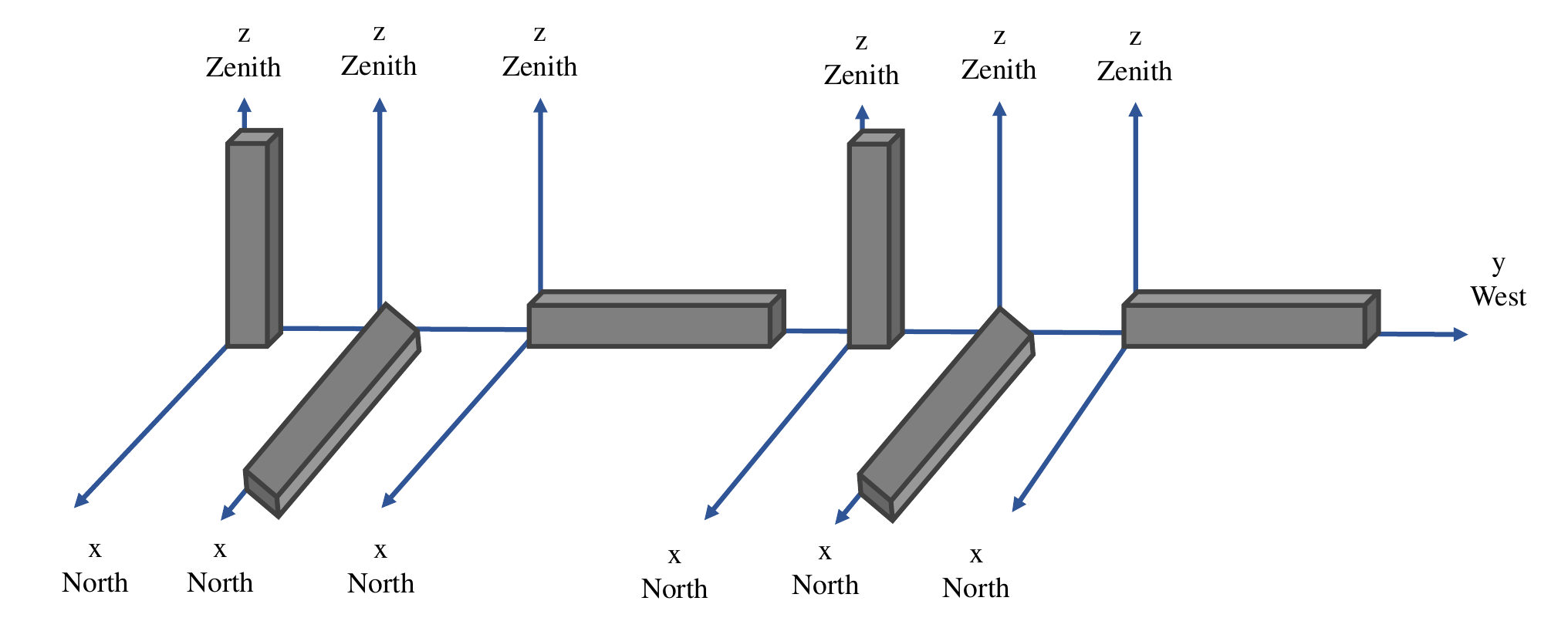}
    \caption{Third setup proposed in this work. Six rectangular cavities oriented towards the three Cartesian axes in the laboratory reference frame thus positioned in the same location.}
    \label{fig:six_oriented_cavities}
\end{figure}

Analyzing this result, it can be seen that the consideration of velocities velocities not only modifies the phase of the voltage but also its amplitude: since velocity variation is sinusoidal along the year, and the voltage amplitude has explicit dependence with the de Broglie wavenumber components, it is expected a sinusoidal variation of the detected signal over the days. In addition, even thought it has been hidden in $\xi$, the resonant behavior is still present in the amplitude of the voltage as we should expect. Finally we can see that the phase has three well-differentiated parts: the first one, which makes reference to the de Broglie wavenumber components and is thus related with the variation of the signal along the year due to velocities; the second one, which is the intrinsic phase of the axion field $\varphi$; and the third, which is made up of only components related to the resonant cavity itself. Special attention is given to the de Broglie phase, where each component of the de Broglie wavenumber is multiplied by the corresponding dimension of the cavity, thus observing that the specific wavenumber component associated with each dimension depends on the orientation of the cavity. This implies that elongating one particular dimension of the cavity potentiates the associated de Broglie wavenumber component.
Thus, in this work three different multi-cavity experiments have been studied: the first one, depicted in Figure \ref{fig:earth_setup}, which is made up of three cavities located in three different latitudes and in the same longitude\footnote{In order to be sensitive to small phase drifts across such long distances, the employment of very precise and stable clocks is mandatory when downconverting the frequencies. For instance, synchronized maser clocks should be used in a realistic experiment.}, oriented towards the same axis direction (i.e., all of them oriented towards Zenith in each laboratory reference frame); the second one, depicted in Figure \ref{fig:three_oriented_cavities}, composed of three cavities positioned in the same location of the Earth but oriented towards different directions, being this setup the practical version of the previous one; and the third one, equal to the latter, but with six cavities, as it is shown in Figure \ref{fig:six_oriented_cavities}. The aim of this study is to employ the phase shift among the cavities induced by the de Broglie wavenumber in order to obtain valuable information about the signal variation along the year. In the first setup, phase shift between cavities is induced by the variation of the velocity components of the axion due to the change of latitude of the laboratory reference frame; in the second one, the phase shift is induced by potentiating different components of the de Broglie wavelength by orienting the cavities towards different directions. As it can be expected, the longer the cavity, the more noticeable will be the effects of this variation. In this way, calculations for these setups have been performed for cavities C1, C2 and C3 of Table \ref{tab:lengths}.

\subsection{Cross-correlation applied to the SNR improvement}

One of the main parts of this work is the study of the SNR improvement by means of cross-correlating the signal from different cavities. For a brief explanation of the cross-correlation and its features, see Appendix \ref{app:cross-correlation_appendix}. A very similar procedure was carried out in \cite{australianos_IEEE}, and here, a summary of the theory involved in the process is provided.

Signal-to-noise ratio is defined as follows:
\begin{equation}
    \mathrm{SNR} = \frac{\mathrm{PSD_{total}}\left(\omega_{a}\right) - \langle \mathrm{PSD}_{\mathrm{noise}}\rangle}{\sigma_{\mathrm{noise}}},
\end{equation}
where $\mathrm{PSD_{total}}$ $=$ $\mathrm{PSD}_{\mathrm{axion}}$ $+$ $\mathrm{PSD_{noise}}$ at the frequency of interest, being $\mathrm{PSD}_{\mathrm{axion}}$ and $\mathrm{PSD_{noise}}$ the power spectral densities of axion and noise, respectively; $\langle \mathrm{PSD_{noise}}\rangle$ is the mean value of the noise PSD;  and $\sigma_{\mathrm{noise}}$ is the noise standard deviation. When considering a single cavity, standard deviation evolves as:
\begin{equation}
    \sigma_{\mathrm{noise}} = \frac{\sigma_{0}}{\sqrt{m}},
\end{equation}
where $\sigma_{0}$ is the standard deviation for only one average, and $m$ is the number of averages considered. This implies that, when increasing the number of averages, the standard deviation decreases, while the mean of the noise level remains constant as a function of time.

When adding more cavities, it has to be remembered that cross-correlations will be performed by pairs, implying that the standard deviation of the noise will evolve as:
\begin{equation}
    \sigma_{\mathrm{noise}} \approx \frac{\sigma_{0}}{2\sqrt{m}}.
\end{equation}

In addition, when considering the total combination of pairs that can be made for a setup of $n$ cavities, the total number of iterations $m$ has to be multiplied by a factor $n\left(n-1\right)/2$. Taking into account these aspects, signal-to-noise ratio for the correlation of $n$ cavities is expected to behave as:
\begin{equation}
    \mathrm{SNR}_{n} \approx 2\sqrt{\frac{n\left(n-1\right)}{2}}\sqrt{m}\, \frac{\mathrm{PSD_{total}\left(\omega_{a}\right)}}{\langle\mathrm{PSD_{noise}}\rangle}.
    \label{eq:SNR_evolution}
\end{equation}
Since the noises from the cavities are completely uncorrelated, both the standard deviation and the mean of the noises will decrease, while the axion power will remain constant.

\section{Numerical results}

In this section we show the simulations results performed in combination with the mentioned theory. It is composed of two parts: the first one, in which cross-correlation technique between different cavities is applied for the improvement of the SNR; and the second one, in which valuable information is extracted from the cross-correlation results in order to study the daily variation of the detected axion signal.

\subsection{Noise in BI-RME3D formulation}

Here, we apply the BI-RME3D method to a specific cavity, in this case the cavity C1. Since the cavity acts like a band-pass filter, the maximum power delivered to the waveguide in the resonant frequency should be the expected noise power, decreasing the value of the power delivered to the waveguide when moving away from the resonance peak.

\begin{figure}[h]
    \centering
    \includegraphics[scale = 0.6]{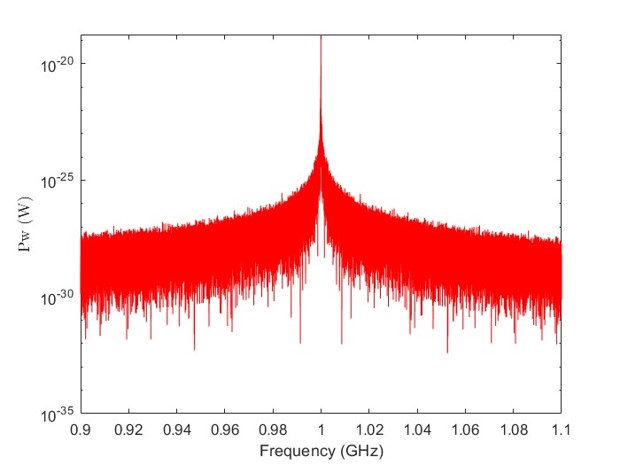}
    \caption{Frequency sweep simulation of extracted power from the port $P_{w}$ for the cavity C1 excited only by noise at a temperature of 4 K.}
    \label{fig:resonant_noise}
\end{figure}

We can now plot a sweep in frequencies of $P_{w}$ in order to show clearly the results obtained in the simulation. Results are depicted in Figure \ref{fig:resonant_noise}, and, as it can be seen, BI-RME3D technique has consistently replicated the expected resonant noise of the cavity.

\subsection{SNR improvement by cross-correlation processing}

In this section we show the SNR improvement by using the cross-correlation processing in a multiple cavity system. The cavity considered in this section is C1, and no kind of phase shift is being considered yet for these calculations. The effects of the phase shifts due to the mechanical tolerances in the manufacturing process are left for future prospects. 

As it was mentioned previously, this study was already performed by \cite{australianos_IEEE}. Nevertheless, the main goal of our study is that, due to the properties of BI-RME3D, it is very straightforward and quick to perform cross-correlation between signals, since BI-RME3D provides the amplitude and phase values of the detected signal. In order to perform the correlation in the Fourier domain, the axion signal and the noise are simulated with values obtained from BI-RME3D method.

\begin{figure}[h]
    \centering
    \includegraphics[scale = 0.7]{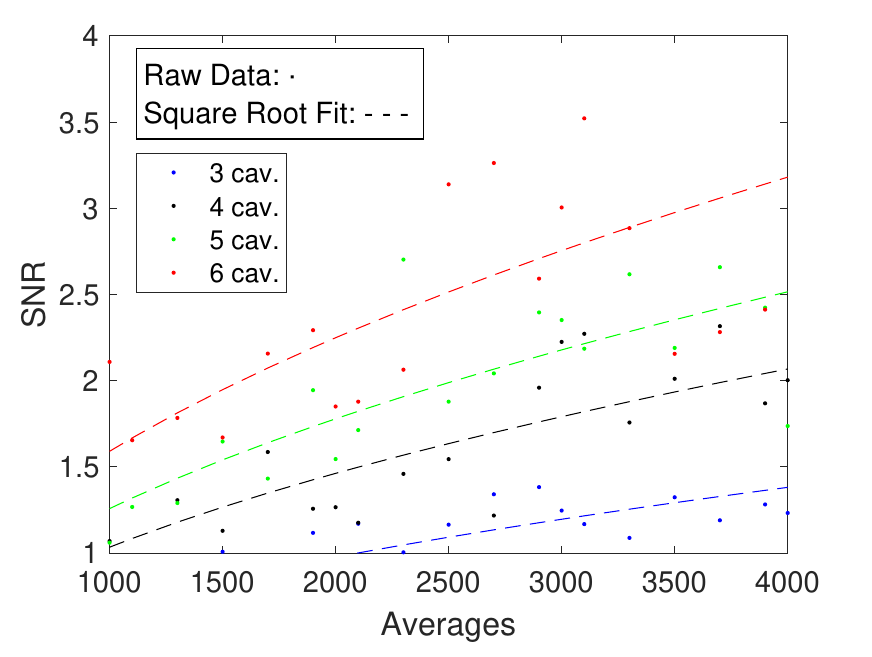}
    \caption{SNR evolution for different number of averages and number of cavities. As it can be seen, data fits well to the expected square root law of SNR (dashed lines) with the number of averages (which is proportional to the exposure time).}
    \label{fig:SNR_improvement}
\end{figure}

Figure \ref{fig:SNR_improvement} illustrates the predicted behaviour of the SNR over the number of averages and cavities in the system, and these results are similar to the ones obtained in \cite{australianos_IEEE}. This method improves the SNR growth rate when comparing to the typical power summation:
\begin{equation}
    \mathrm{SNR_{n}} = n\sqrt{m}\,\frac{\mathrm{PSD_{total}\left(\omega_{a}\right)}}{\langle\mathrm{PSD_{noise}}\rangle}.
\end{equation}

The improvement of the cross-correlation method compared with the summation one depends on the number of cavities considered and it can be obtained as follows:
\begin{equation}
    \frac{\mathrm{SNR_{\star}}}{\mathrm{SNR_{+}}} = \sqrt{2}\,\sqrt{1 - \frac{1}{n}},
\end{equation}
where $\mathrm{SNR_{\star}}$ and $\mathrm{SNR_{+}}$ are the $\mathrm{SNR}$ in cross-correlation and summation cases, respectively. Results for different number of cavities are shown in Table \ref{tab:improvement}.

\begin{table}[htbp]
\centering
\begin{tabular}{c | c}
\hline
Number of cavities $n$ & Improvement Ratio $\mathrm{SNR_{\star}}/\mathrm{SNR_{+}}$\\
\hline
2 & 1.00\\
3 & 1.15\\
4 & 1.22\\
5 & 1.26\\
6 & 1.29\\
\hline
\end{tabular}
\caption{Improvement ratio between the cross-correlation and summation cases.\label{tab:improvement}}
\end{table}

\begin{figure}
    \centering
    \includegraphics[scale = 0.7]{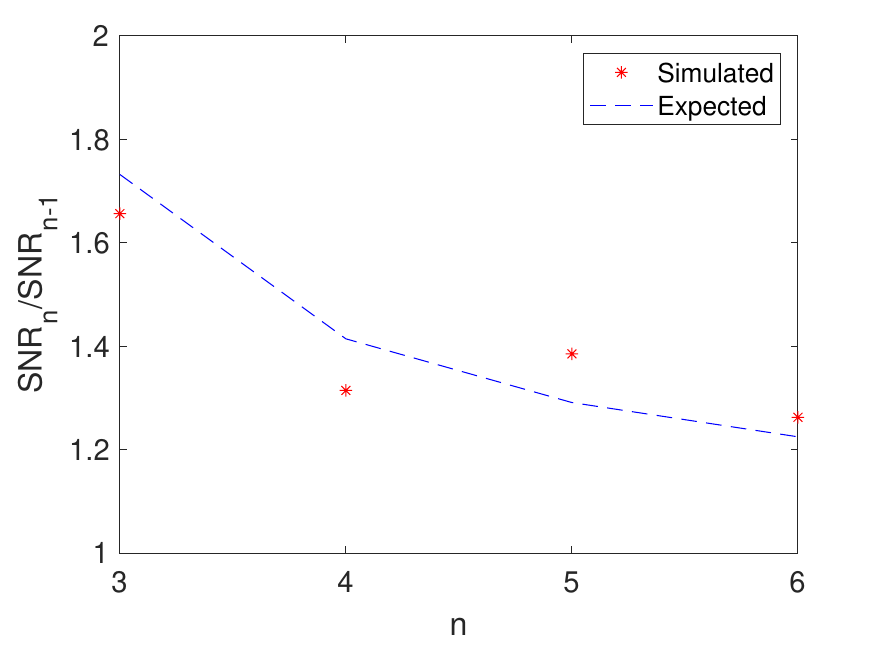}
    \caption{SNR behavior with the number of cavities. Ratio between the SNR for $n$ cavities case and $n-1$ cavities case is shown as a function of the number of cavities. An asymptotic behavior is observed as expected from Eq.\ (\ref{eq:SNR_behaviour_cavities}) when increasing $n$. It can be observed that the obtained results fit well to the theoretical expression.}
    \label{fig:SNR_cavities}
\end{figure}

Complementing this study, we have studied the evolution of the SNR with the number of cavities. If the behavior followed by the SNR is described by Eq.\ (\ref{eq:SNR_evolution}), it implies that the ratio between the case with $n$ cavities and $n-1$ cavities should be:
\begin{equation}
    \frac{\mathrm{SNR}_{n}}{\mathrm{SNR}_{n-1}} = \sqrt{\frac{n\left(n-1\right)}{\left(n-1\right)\left(n-2\right)}}.
    \label{eq:SNR_behaviour_cavities}
\end{equation}
Results of this calculation are shown in Figure \ref{fig:SNR_cavities}, observing a good agreement with the expected behavior. In addition, this explains why an asymptotic behavior is observed when increasing the number of cavities: if $n$ tends to infinity the limit tends to $1$, just concluding that a system with a large number of cavities will not observe a noticeable improvement when introducing one additional cavity, although the improvement is significant when having a reduced number of cavities.

To conclude this section, the BI-RME3D theory has allowed to study the cross-correlation between the cavities with realistic values of amplitude and phase of the detected signals for axion and resonant noise. Moreover, it is capable of reproducing the results from the bibliography about the SNR evolution with the number of averages and cavities, and it is able to prove the asymptotic behavior of the SNR with the number of cavities added to the experimental setup. 

\subsection{Directionality and daily modulation of the axion signal}

In this section, we perform numerical calculations of the effect of directionality in an axion experiment, and we also calculate the cross-correlation between the signal from different cavities of an experimental testbed. For such purpose, Eq.\ (\ref{eq:V_c_final}) has to be considered. As we mentioned previously, the effect of considering velocities introduces phase-shifts between the different cavities in our setup. 

\begin{figure}[h]
    \centering
    \includegraphics[scale = 0.35]{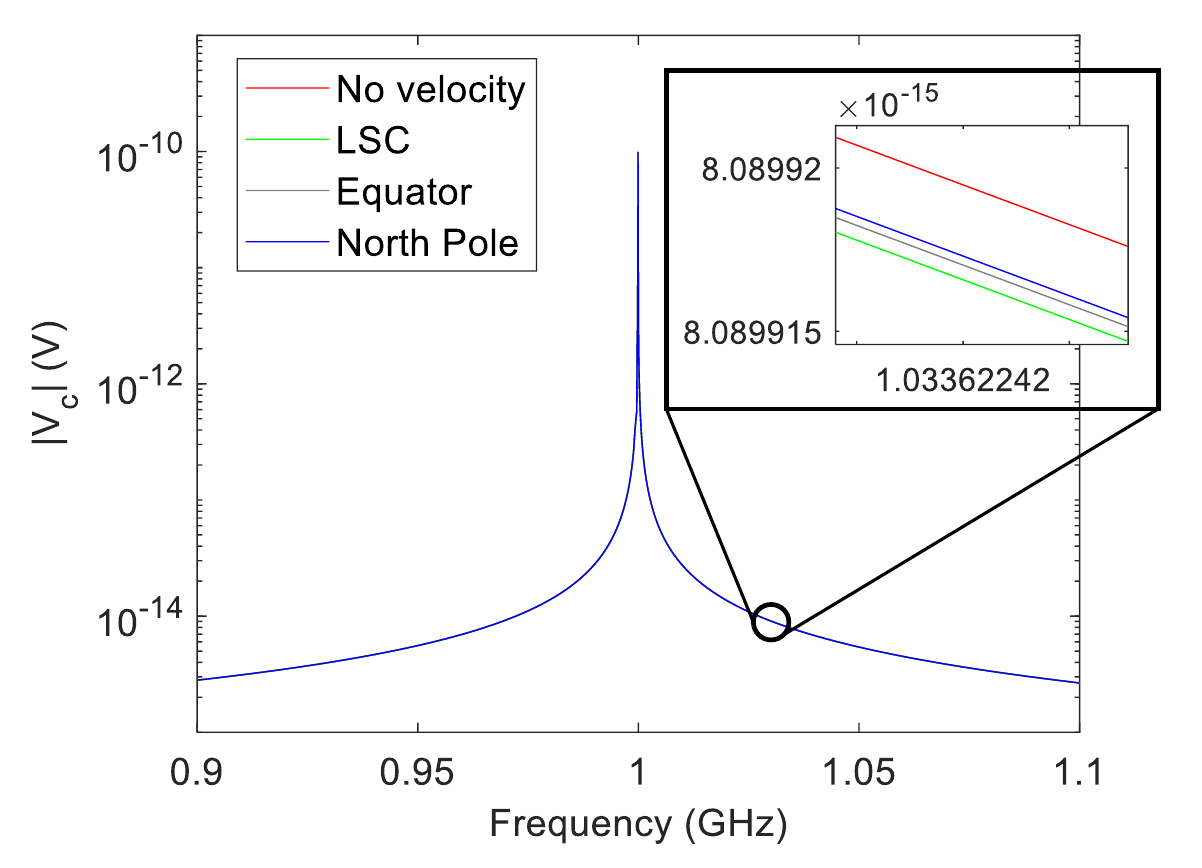}
    \includegraphics[scale = 0.35]{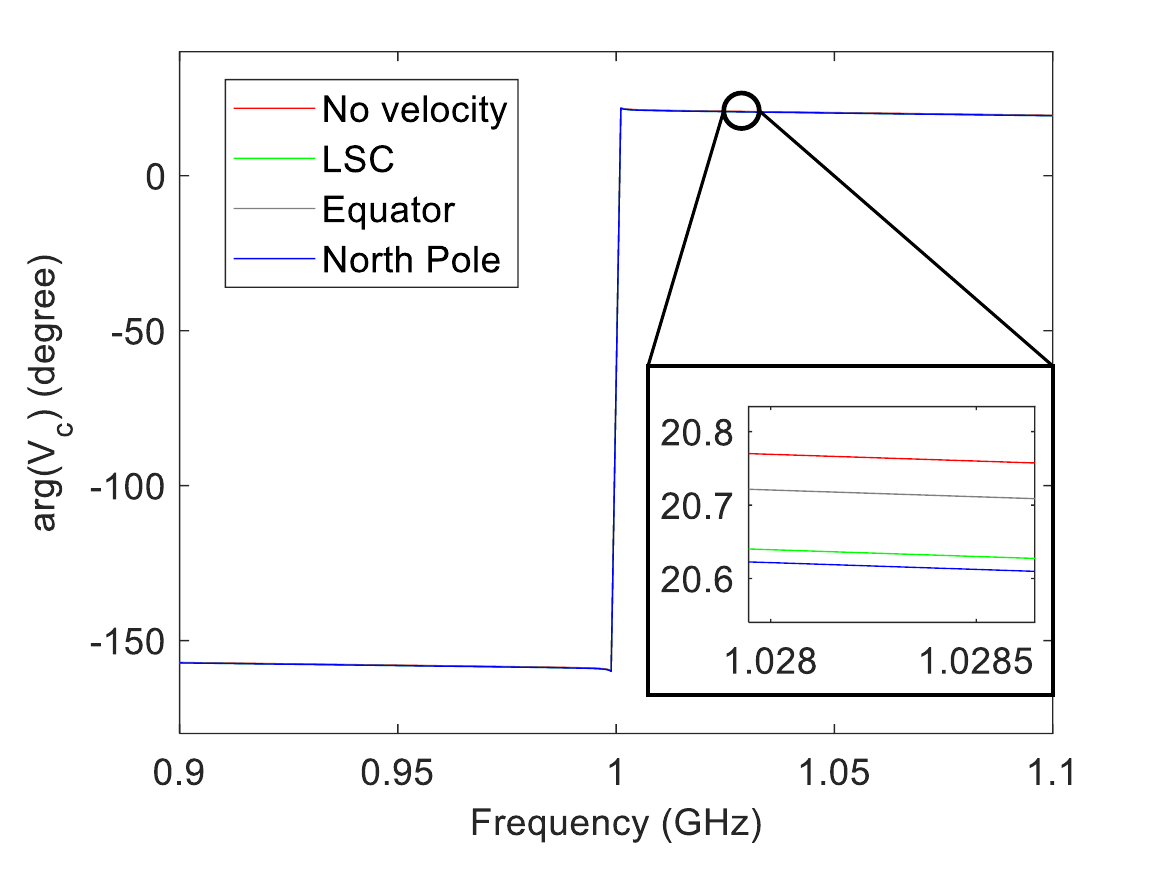}
    \caption{Values of the detected voltage $V_{c}$ for the C1 cavity in three different locations. A frequency sweep was made around the resonance of the mode of interest. Left: Amplitude of the detected voltage. Right: Phase of the detected voltage.}
    \label{fig:setup_1_C1}
\end{figure}

\begin{figure}
    \centering
    \includegraphics[scale = 0.35]{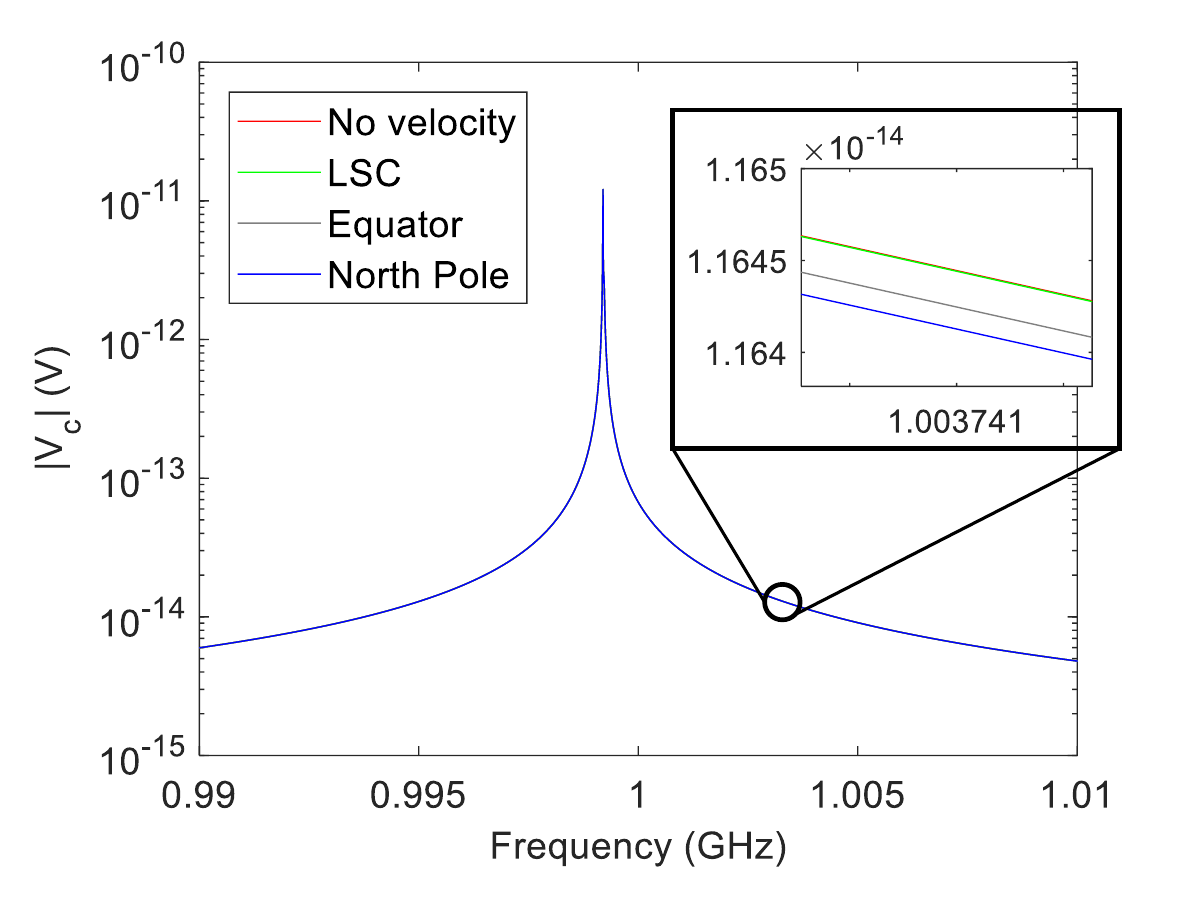}
    \includegraphics[scale = 0.35]{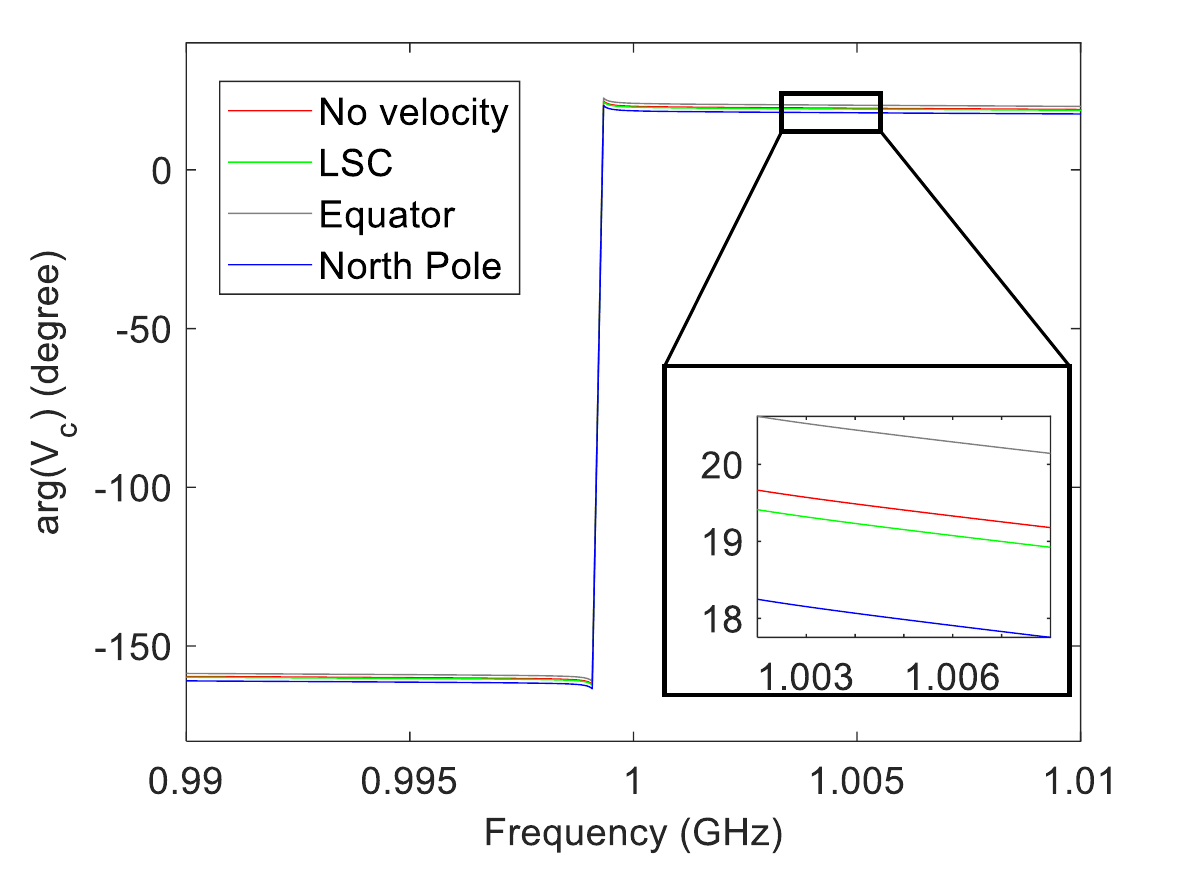}
    \caption{Values of the detected voltage $V_{c}$ for the C3 cavity in three different locations. A frequency sweep was made around the resonance of the mode of interest. Left: Amplitude of the detected voltage. Right: Phase of the detected voltage.}
    \label{fig:setup_1_C3}
\end{figure}

\begin{figure}
    \centering
    \includegraphics[scale = 0.5]{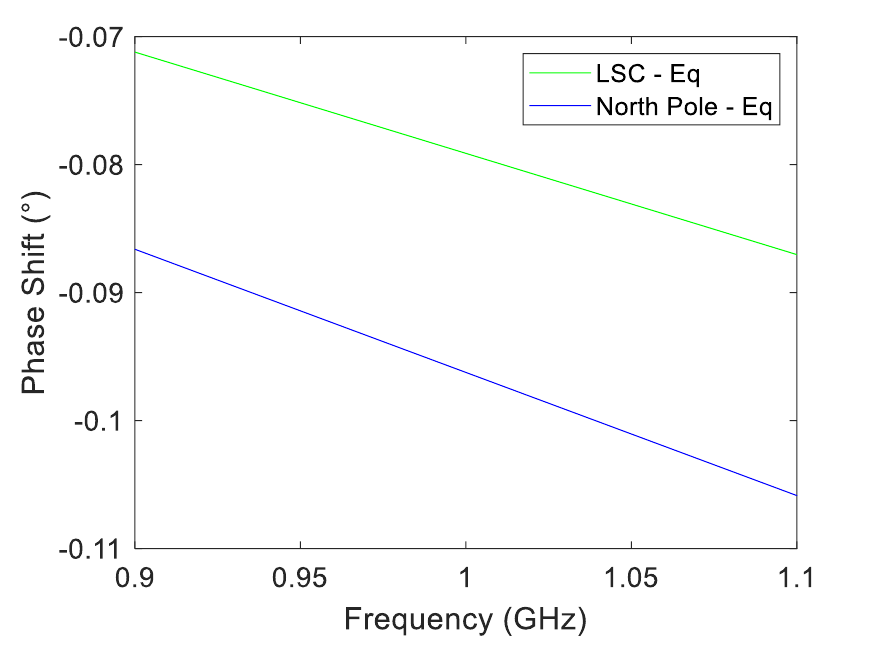}
    \includegraphics[scale = 0.5]{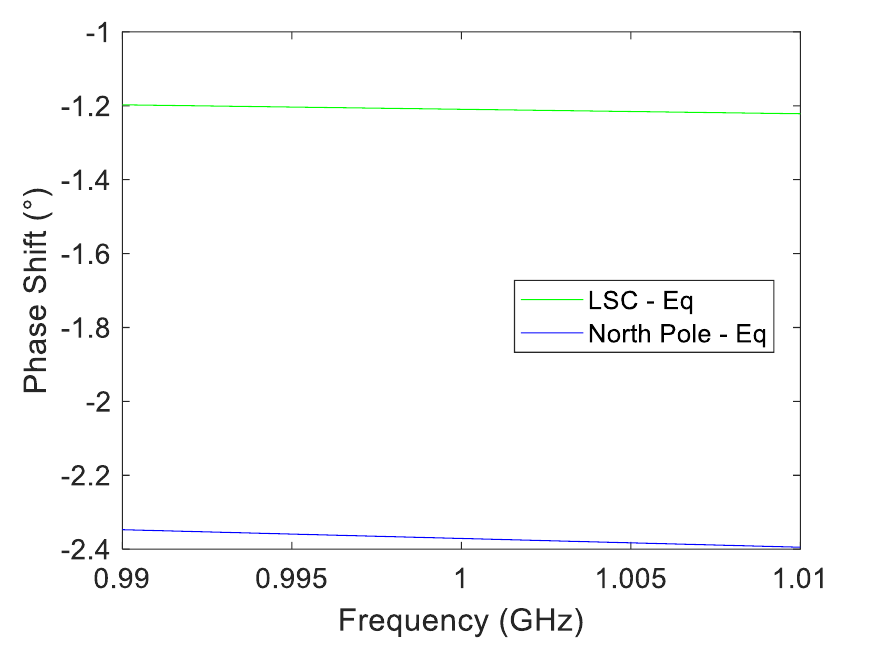}
    \caption{Phase shifts of the recorded voltages $V_{c}$ for cavities C1 (left) and C3 (right) between two pairs of cavities of Setup 1. Green line makes reference to the phase shift between LSC cavity and Equator cavity, and the blue line is related to the phase shift between North Pole cavity and Equator cavity.}
    \label{fig:phase_shifts_C1_C3}
\end{figure}

Focusing first on Setup 1 (Figure \ref{fig:earth_setup}), numerical results can be seen in Figures \ref{fig:setup_1_C1} and \ref{fig:setup_1_C3} for cavities C1 and C3, respectively (C2 has been omitted since the results obtained are an intermediate step between C1 and C3). These results were obtained for day 1 of the year in three different latitudes. As it can be seen, the consideration of velocities in the detected voltage modifies both amplitude and phase. The magnitude of this modification depends on the length of the considered cavity. How much this affects to the results can be easily measured with the phase shift between two of the setup cavities (see Figure \ref{fig:phase_shifts_C1_C3}). As it is expected, phase shifts obtained for cavity C3 are more than one order of magnitude higher than for C1. Even though the phase shifts for C1 are hardly distinguishable, the C3 ones could be measured in a real experiment with the appropriate technology. 

The value of these phase shifts is directly related to the velocity components, which are shown in Table \ref{tab:v_components}: wider differences between velocity components in the studied latitudes result in higher phase shifts between cavities, as it is expected. More specifically, when considering long cavities as in the case of C3, the velocity component that governs the phase shift is the one related to the axis toward the cavity is oriented. In our case, since cavities are oriented towards $z$ axis (Zenith direction in laboratory reference frame), the phase shift between North Pole and Equator in C3 case is the strongest one since the difference between both velocity components is maximum.

\begin{table}[htbp]
\centering
\begin{tabular}{c | c | c | c}
\hline
Location & $v_{x}$ $\left(\cdot 10^{5}\ \mathrm{m/s}\right)$ & $v_{y}$ $\left(\cdot 10^{5}\ \mathrm{m/s}\right)$ & $v_{z}$ $\left(\cdot 10^{5} \ \mathrm{m/s}\right)$\\
\hline
LSC & 2.141 & 0.559 & 0.168 \\
Equator & 1.686 & 0.558 & -1.330\\
North Pole & 1.330 & 0.563 & 1.686\\
\hline
\end{tabular}
\caption{Velocity components in the three different latitudes studied for the day 1 of the year in the laboratory reference frame.\label{tab:v_components}}
\end{table}

\begin{figure}[h]
    \centering
    \includegraphics[scale = 0.6]{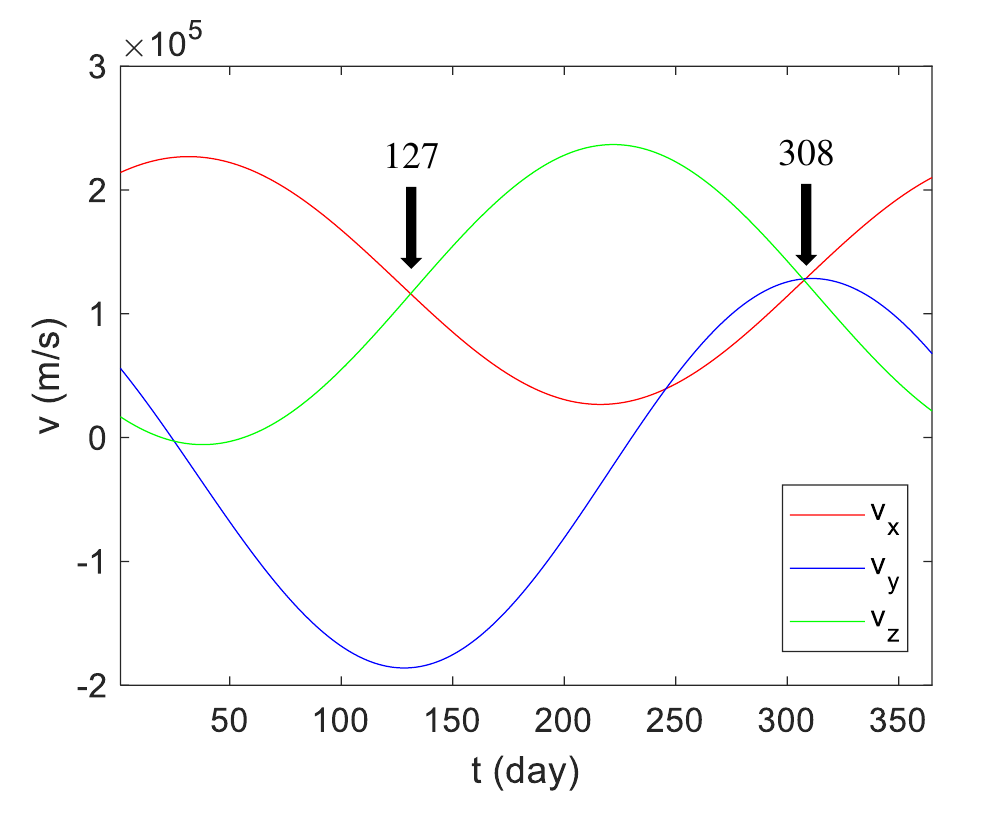}
    \caption{Velocity components variation along the year in the LSC laboratory reference frame. Days 127 and 308 are pointed out.}
    \label{fig:v_components_LSC}
\end{figure}

The disadvantage of this setup is clear because it can be cumbersome to synchronize three cavities in three different locations of the Earth. Related to the logistics problem, the clocks for each cavity need to be different, so it would be even more challenging to keep them well synchronized (it is common to have slow drifts, even with atomic clocks). Thus, it is natural to ask if similar results could be obtained with an easier-to-handle setup, like Setup 2. Since phase shift is only related to differences between velocity components, if Setup 2 is located in the appropriate latitude and day of the year, comparable results should be obtained. For that, first give a look to the velocity components along the year in the LSC location (see Figure \ref{fig:v_components_LSC}).

\begin{figure}
    \centering
    \includegraphics[scale = 0.51]{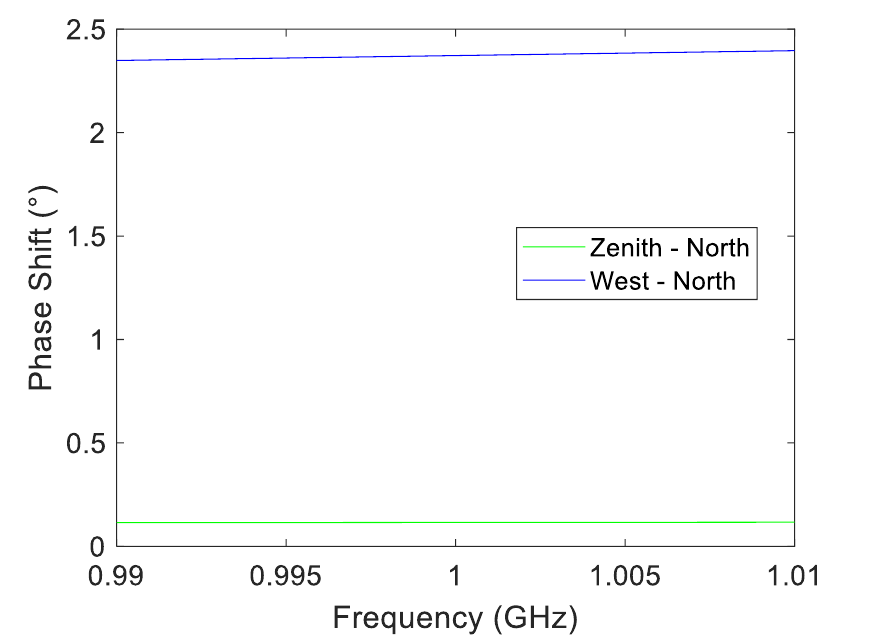}
    \includegraphics[scale = 0.51]{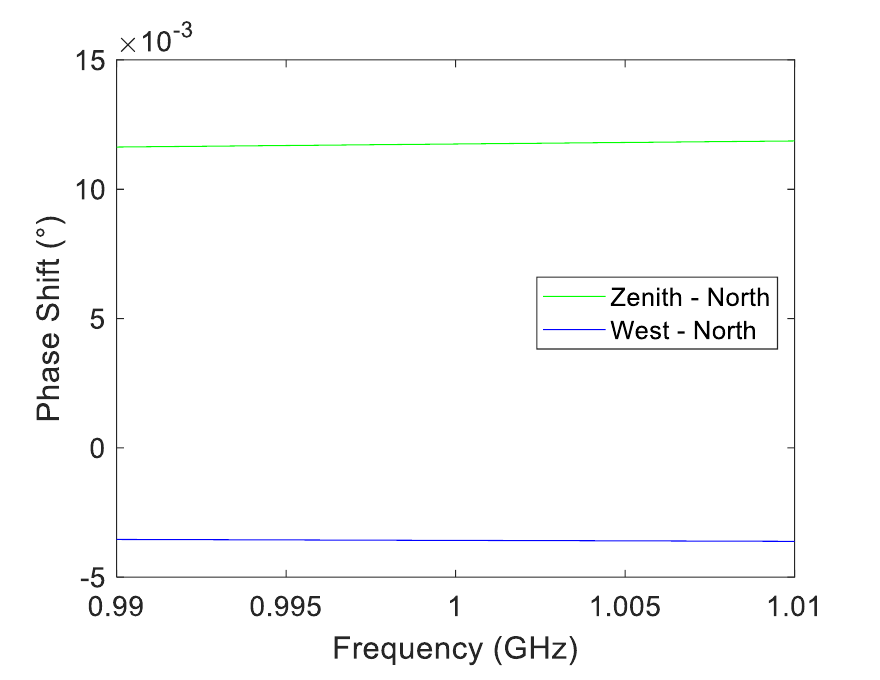}
    \caption{Phase shifts of the recorded voltage $V_{c}$ obtained for Setup 2 in days 127 (left) and 308 (right) of the year. In green, phase shift between the pair of cavities oriented towards Zenith ($z$ axis) and North ($x$ axis); in blue, phase shift between the pair of cavities oriented towards West ($y$ axis) and North ($x$ axis).}
    \label{fig:phase_shifts_setup_2}
\end{figure}

Maximum difference between two of the velocity components is accomplished in the day 127 of the year, and the three velocities are approximately equal in the day 308. Thus, when Setup 2 is considered in the day 127, similar phase shifts to the ones in Setup 1 are expected when studying the phase shift pairs West-North and West-Zenith, as well as a low one when studying Zenith-North directions. On the other hand, low phase shifts can be obtained when looking into the day 308 for every considered pair. Results for C3 cavity are shown in Figure \ref{fig:phase_shifts_setup_2}, supporting the mentioned expectations.

The consideration of velocities grants the experiment with directional sensitivity under appropriate conditions. In order to exploit this effect even more, we can recover the cross-correlation technique: since the cross-correlation of two signals in Fourier domain is directly proportional to the phase shift between them, the technique explained along this work can be applied to this directional-sensitive experiment (see Appendix \ref{app:cross-correlation_appendix}). This can be demonstrated by assuming two sinusoidal signals with the same amplitude and frequency but with different initial phase. When calculating the cross-correlation in Fourier domain, the result is:
\begin{equation}
    \mathrm{FT}\left[h\right] = \frac{A^2}{4}\left[e^{-i\left(\varphi_{1} - \varphi_{2}\right)}\delta\left(\omega - \omega_{0}\right) + e^{-i\left(\varphi_{2} - \varphi_{1}\right)}\delta\left(\omega + \omega_{0}\right)\right],
\end{equation}
where $h$ is the cross-correlation, $A$ is the signal amplitude, $\varphi_{1}$ and $\varphi_{2}$ their respective initial phases, and $\omega_{0}$ the signal angular frequency; $\mathrm{FT}$ represents the standard Fourier Transform, as it has been defined in Appendix \ref{app:cross-correlation_appendix}.

\begin{figure}[b]
    \centering
    \includegraphics[scale = 0.5]{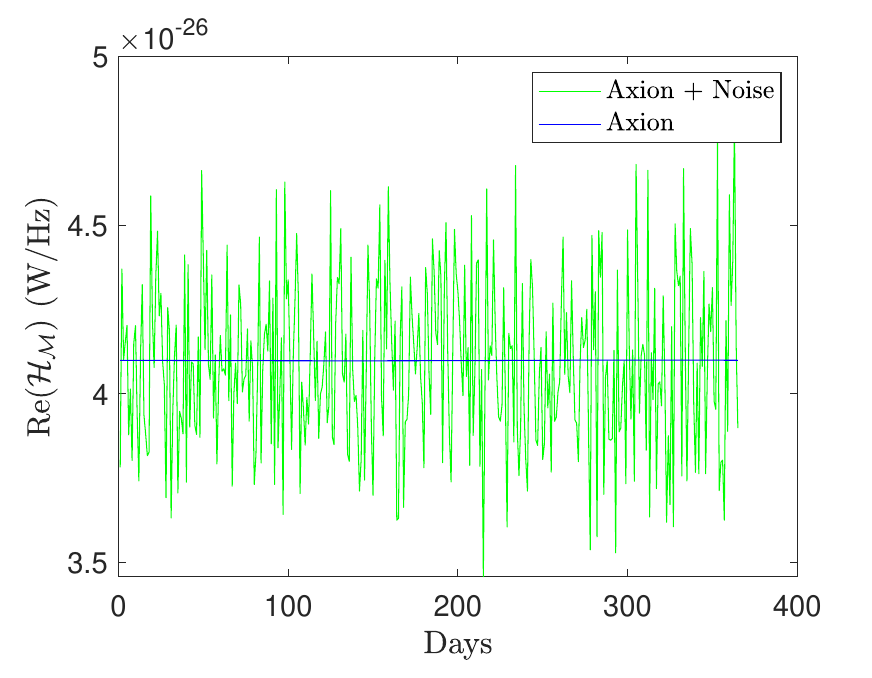}
    \includegraphics[scale = 0.5]{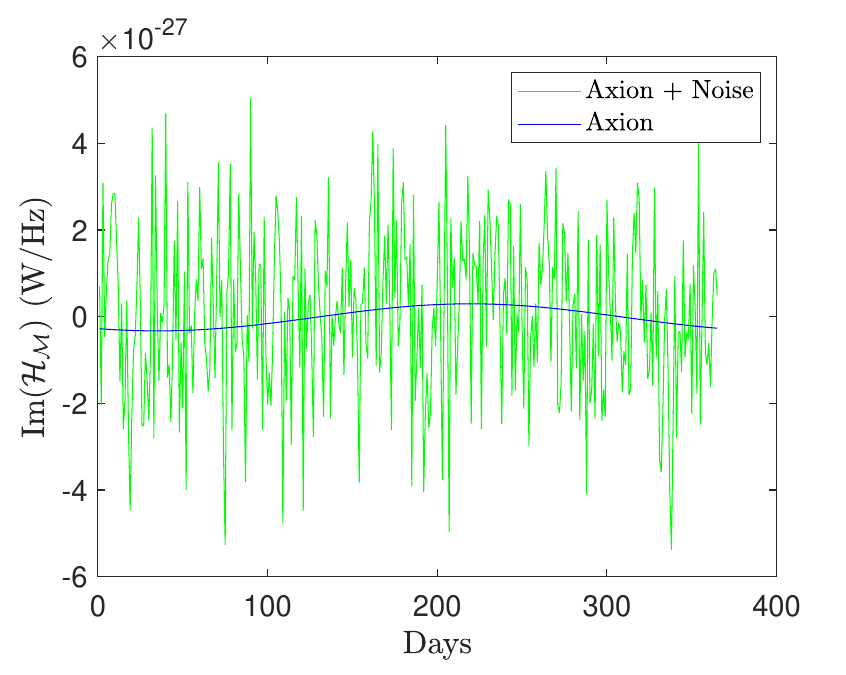}
    \caption{Cross-correlation maximum variation along the year for Setup 3 made up of C3 cavities for SNR = 16. Real (left) and imaginary (right) parts are shown. Green curve makes reference to noisy data, considering both axion and resonant noise, while the blue curve makes only reference to the axion.}
    \label{fig:daily_variation_C1}
\end{figure}

\begin{figure}
    \centering
    \includegraphics[scale = 0.5]{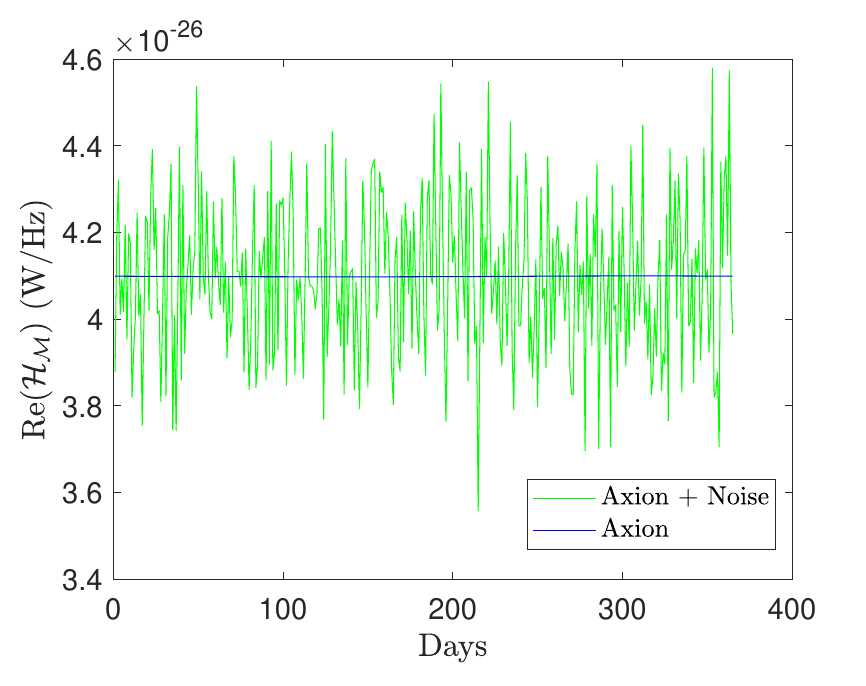}
    \includegraphics[scale = 0.5]{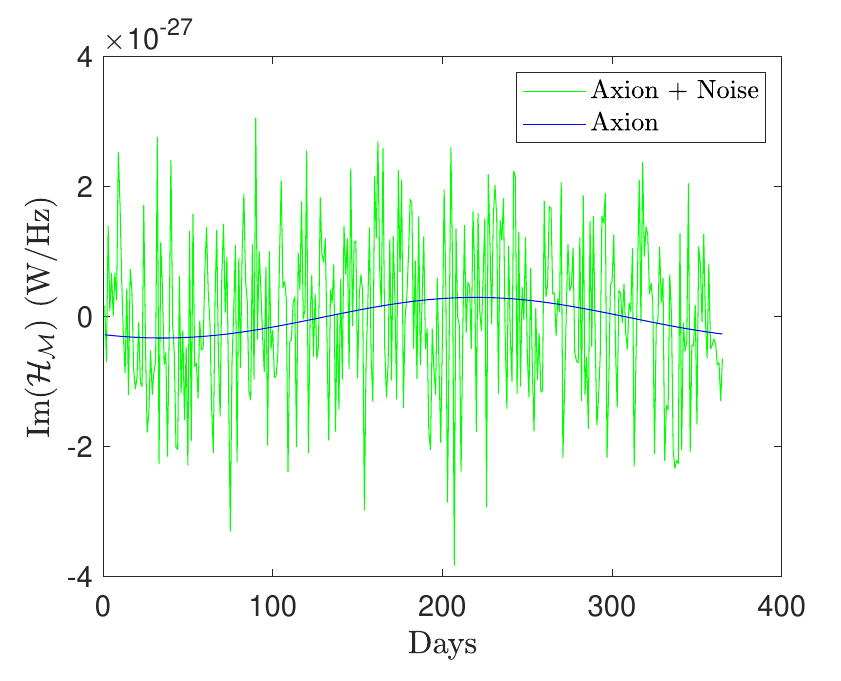}
    \caption{Cross-correlation maximum variation along the year for Setup 3 made up of C3 cavities for SNR = 23. Real (left) and imaginary (right) parts are shown. Green curve makes reference to noisy data, considering both axion and resonant noise, while the blue curve makes only reference to the axion.}
    \label{fig:daily_variation_C3}
\end{figure}

The aim of this study is to observe how strong is the variation of the cross-correlation result induced by the axion signal modulation. Results are obtained for setup 3 and C3 cavity since the phase shift is higher in this case. Results are shown in Figures \ref{fig:daily_variation_C1} and \ref{fig:daily_variation_C3}, for two different values of SNR: SNR $= 16$ and SNR $= 23$. To obtain these results, the maximum of the cross-correlated spectrum $\mathcal{H}$ in resonance has been extracted each day. We denote this criteria by putting a subindex $\mathcal{M}$ in the magnitudes that are calculated from the cross-correlated spectrum maximums along the year. In this way, $\mathrm{Re\left(\mathcal{H}_{\mathcal{M}}\right)}$ makes reference to the real part of the spectrum maximum for each day, and $\mathrm{Im\left(\mathcal{H}_{\mathcal{M}}\right)}$ is the same but for the imaginary part. This is one of the most important results of this work, since it is an actual observable in which we see the daily variation along the year of the cross-correlation signal induced by the variation of the phase shift between cavities due to the effect of velocities measured in the laboratory reference frame. 

In order to characterize the daily modulation, a ratio $R$ has been defined as follows,

\begin{equation}
    R = \frac{\mathrm{max\{Im\left(\mathcal{H}_{axion,\,\mathcal{M}}\right)\}} - \mathrm{min\{Im\left(\mathcal{H}_{axion,\,\mathcal{M}}\right)\}}}{\sigma\{\mathrm{Im}\left(\mathrm{\mathcal{H}_{noise,\,\mathcal{M}}}\right)\}},
\end{equation}
where the numerator is the difference between the maximum and the minimum of $\mathrm{Im\left(\mathcal{H}_{axion,\,\mathcal{M}}\right)}$ of the whole year, and the denominator is the standard deviation of $\mathrm{Im\left(\mathcal{H}_{noise,\,\mathcal{M}}\right)}$. In this way, this ratio can be seen as a new kind of SNR but for the modulation case. For both SNR studied, the $R$ ratios are $R = 0.33$ and $R = 0.51$, respectively. In order to observe a noticeable modulation, high SNR has been considered.

Another aspect is that the variation along the year is stronger when looking to the imaginary part of the signal than when looking to the real part (where the oscillation cannot be noticeable). Since the real part of the correlation in Fourier domain evolves like a cosine of the phase shift between cavities while the imaginary part evolves as a sine, it implies that in the approximation of low phase shifts (as the ones studied here) the perturbations to the real part go as a second order perturbation while in the imaginary part is a first order perturbation. As a result, it is more advantageous to look to the imaginary part, since any change in phase shift will be remarkably more noticeable. In addition, as it can be observed in the results, the real part of the cross-correlation adopts higher values than the imaginary part. This can be also explained with the low angle approximation, since the imaginary part is proportional to the angle while the real part is proportional to one minus the angle square over two. This makes that, provided that the angle is low, the real part will always adopt higher values than the imaginary part.

Observing the results obtained, we can conclude that the provided method for observing the daily modulation of the axion signal can be employed for the characterization of the velocity distribution (post-detection information about the axion), since any deviation on the assumed velocities from the real ones will be noticeable when looking to the daily modulation, since it would be different. However, provided the mentioned phase shift, this modulation is not high enough to be employed as an indirect detection method, since high SNR must be obtained in order to notice this daily modulation. For this modulation to be more significant, one should achieve higher phase shifts between cavities, which could be reached by considering longer cavities. In this way, with higher phase shifts the daily modulation would be more significant hence increasing $R$ for a given SNR, just allowing for this method to be employed as an indirect detection technique in the case when $R$ is greater than SNR. 

An important aspect to consider is whether the loss of form factor, when increasing the mode order $l$ (to extend the cavity length while maintaining the resonant frequency unchanged), is worthwhile in order to observe the directionality effect. It is well known that when increasing the $l$ index the form factor of a rectangular resonator will decrease as reported by this simple equation \cite{directional_axion_detection},
\begin{equation}
   C = \frac{64}{\pi^{4}l^{2}}. 
\end{equation}
As a consequence, when we enlarge the third cavity dimension the form factor will be reduced according to this equation. Now we are going to analyze this effect for the C1 and C3 cavities.
The decrease in this parameter when increasing $l$ is
\begin{equation}
    \frac{C_{3}}{C_{1}} = \frac{1}{441} \approx 0.002,
\end{equation}
where $C_{1}$ and $C_{3}$ are the form factors of cavities C1 and C3, respectively. The increment in volume is
\begin{equation}
    \frac{\mathcal{V}_3}{\mathcal{V}_1} = 20.8,
\end{equation}
where $\mathcal{V}_{1}$ and $\mathcal{V}_{3}$ are the volumes of cavities C1 and C3, respectively. The increment in quality factor is
\begin{equation}
    \frac{Q_{3}}{Q_{1}} \approx 1.6,
\end{equation}
where $Q_{1}$ and $Q_{3}$ make reference to the unloaded quality factors of cavities C1 and C3, respectively. Hence the detected power decreases approximately by a factor of 15. In order to solve this loss, the same study performed in this work could have been made with a cylindrical cavity supporting $\mathrm{TM}_{010}$, whose form factor does not depend on the cavity length, making it perfect for this kind of study\footnote{However, one should be careful with mode clustering when increasing the length of the cavity, since it may hinder the axion mode identification if another resonant mode is close enough.}. In addition, experiments based on the use of multicavities could also be employed for the analysis of directional sensitivity, improving the configuration proposed here \cite{RADES_1, RADES_2}. 

\section{Conclusions}

In this work, the study of multiple-cavity setups has been developed in order to improve the evolution of the signal-to-noise ratio when increasing the integration time as well as the interferometric analysis of the effect of directionality on the detection of dark matter axions.\\ 

With this aim, the BI-RME3D method has been used, allowing to relate the problem of the axion-photon decay inside a resonant cavity with the classical electromagnetic microwave network theory. This formulation not only provides the detected power extracted from the cavity excited by a potential axion-photon conversion, but also yields the phase of the signal voltage measured in the coupled port. In order to perform a realistic simulation, the resonant noise extracted from the cavity has been simulated through BI-RME3D formulation.\\

Cross-correlation between potential signals from a multiple-cavity setup has been calculated. SNR after the cross-correlation presented a higher growth rate with respect to the exposure time by a factor $\sqrt{2}\sqrt{1 - (1/n)}$ in comparison with the power-summation method, making it clear that cross-correlation could be potentially employed in a real setup for the improvement of the exposure time of a realistic experiment. For instance, a six cavities setup shows an improvement close to 30\% in SNR when comparing to power-summation method. This growth rate has been shown to increase asymptotically with the number of cavities.\\

With the aim of studying the directionality effect in axion detection, three different setups were proposed. BI-RME3D formulation provides both the amplitude and the phase of the detected voltage, so the phase shift between the cavities of the setups has been properly computed. When performing this calculation for the longest cavities considered, signals measured from different cavities differ in their phase in more than $2^{\circ}$. This phase shift could be strong enough to be noticed in a real experiment, and could reach higher values when considering longer cavities. Taking profit from this phase shift, cross-correlation was calculated for setup 3, and a daily modulation in the cross-correlation was observed due to the variation of the phase shifts between cavities generated by the change of the laboratory reference frame velocity across the year. This modulation is considerably more noticeable in the imaginary part of the cross-correlation signal, since it is directly proportional to the phase shift variation. A ratio has been calculated in order to compare this modulation with the thermal noise, obtaining values of $R=0.33$ and $R=0.51$ for two high SNR considered. Although not a dominant effect, it has been proved that daily modulation in a directional setup is likely to be observed (provided that a high SNR is reached), and it can be used to characterize the velocity distribution of the axion field. Finally, an estimation of the power loss due to the increase in mode order has been carried out, revealing that, in order to observe the directionality effect in the C3 cavity (the longest one), the detected power decreases by a factor of 15 compared to that of the C1 cavity. This issue is discussed in this work, highlighting that it could be solved either by employing a cylindrical cavity supporting the $\mathrm{TM}_{010}$ mode, or by using a multi-filter setup, where the final cavity is composed by individual cavities concatenated by irises.

\appendix
\section{Frequency distribution function of the movable axion}
\label{app:MB_distribution}
In this section, we are going to deduce the expression (\ref{eq:MB_distribution}) and we will follow a similar approach as in \cite{Maxwell_Boltzman_distrib}.

As it was mentioned previously, dark matter axions will follow a Maxwell-Boltzmann distribution in the Milky Way (MW) reference frame:
\begin{equation}
    fd^{3}v = \left[\frac{\beta}{\pi}\right]^{3/2}e^{-\beta v^2}d^{3}v,
\end{equation}
where $f$ is the probability density function, $d^{3}v$ $=$ $dv_{x}\,dv_{y}\,dv_{z}$ and $\beta$ $=$ $m_{a}/\left(2k_{B}T\right)$. In order to obtain the distribution in the laboratory reference frame, we should define three velocities that participate in the process:
\begin{equation*}
    v \equiv \mathrm{axion\, velocity\, in\, MW\, frame},
\end{equation*}
\begin{equation*}
    v_{l} \equiv \mathrm{axion\, velocity\, in\, laboratory\, frame},
\end{equation*}
\begin{equation*}
    v_{lab} \equiv \mathrm{laboratory\, reference\, frame\, velocity\, with\, respect\, to\, the\, center\, of\, the\, MW}.
\end{equation*}

Following this, these three velocities fulfill the classical condition $\vec{v}_{l}$ $=$ $\vec{v}$ $-$ $\vec{v}_{lab}$. Let us now consider that $\vec{v}_{lab}$ changes slowly with time, so we can write:
\begin{equation}
    d^{3}v_{l}\approx d^{3}v; \ \ \ \ \ \ \ \ \ \ d^{3}v_{l} = dv_{l,x}\,dv_{l,y}\,dv_{l,z},
\end{equation}
implying that
\begin{equation}
    f\, d^{3}v \approx f\, d^{3}v_{l} = \left[\frac{\beta}{\pi}\right]^{3/2}e^{-\beta v^{2}}d^{3}v_{l}.
\end{equation}
Proceeding with the angular integration:
\begin{equation}
    \int_{0}^{2\pi}\int_{0}^{\pi}f\, v_{l}^2\,\sin\theta\, d\theta\, d\phi\, dv_{l} = \int_{0}^{2\pi}\int_{0}^{\pi}\left[\frac{\beta}{\pi}\right]^{3/2}e^{-\beta v^{2}}v_{l}^2\,\sin\theta\, d\theta\, d\phi\, dv_{l},
\end{equation}
now we can rewrite both sides of the expression as follows:
\begin{equation}
    F\, dv_{l} = \left[\frac{\beta}{\pi}\right]^{3/2}v_{l}^2\int_{0}^{2\pi}\int_{0}^{\pi} e^{-\beta\left(\vec{v}_{l}\, +\, \vec{v}_{lab}\right)^2}\sin\theta\, d\theta\, d\phi\, d v_{l},
\end{equation}
where $F$ has been defined as
\begin{equation}
    F \equiv \int_{0}^{2\pi}\int_{0}^{\pi}f\, v_{l}^2\,\sin\theta\, d\theta\, d\phi.
\end{equation}
Let us now develop the dot product between $\vec{v}_{l}$ and $\vec{v}_{lab}$:
\begin{equation}
    \vec{v}_{l}\cdot \vec{v}_{lab} = v_{l}\,\hat{u}_{r}\cdot \vec{v}_{lab} = v_{l}\left(v_{lab,x}\,\sin\theta\,\cos\phi + v_{lab,y}\,\sin\theta\,\sin\phi + v_{lab,z}\,\cos\theta\right).
\end{equation}
Since an integration over a sphere is being made, we can freely choose the direction of the integration axes. In particular, the $z$ axis can be selected to be orientated towards $\vec{v}_{lab}$. Following this, the result obtained is:
\begin{equation}
\begin{aligned}
    F\, dv_{l} &=  \left[\frac{\beta}{\pi}\right]^{3/2} v_{l}^2\int_{0}^{2\pi}\int_{0}^{\pi}e^{-\beta\left(v_{l}^2\, +\, v_{lab}^{2}\, +\, 2\,v_{l}v_{lab}\cos\theta\right)}\,\sin\theta\, d\theta\, d\phi\, dv_{l}\\
    & = \left[\frac{\beta}{\pi}\right]^{3/2}v_{lab}^{2}\,e^{-\beta v_{l}^2\, -\, \beta v_{lab}^2}\,\left[\frac{2\pi \sinh\left(2\, v_{l}\,v_{lab}\, \beta\right)}{v_{l}\,v_{lab}\,\beta}\right]\,d v_{l}\\
    & = 2\left[\frac{\beta}{\pi}\right]^{1/2}\frac{v_{l}}{v_{lab}}\, e^{-\beta v_{l}^{2}\, -\, \beta v_{lab}^{2}}\, \sinh\left(2\,v_{l}\,v_{lab}\,\beta\right)\,d v_{l}.
\end{aligned}
\end{equation}
The independent variable can be now changed into frequency taking into account that $E$ $=$ $\frac{1}{2}m_{a}v_{l}^2$, which is the kinetic energy of axions in the laboratory frame:
\begin{equation}
    f\left(E\right)dE = 2\left[\frac{\beta}{\pi}\right]^{1/2}\frac{dE}{m_{a}\,v_{lab}}e^{-\beta v_{lab}^2\, -\, 2\beta E/m_{a}}\sinh\left(2\beta\, v_{lab} \sqrt{\frac{2E}{m_{a}}}\right),
\end{equation}
and now the theorem of conservation of probabilities when changing variables can be applied
\begin{equation}
    f\left(E\right) dE = f\left(\nu\right)d\nu \Rightarrow f\left(\nu\right) = f\left(E\right)\frac{dE}{d\nu} = \frac{f\left(E\right)}{d\nu/dE}.
\end{equation}
The expression that relates energy with frequency is
\begin{equation}
    h\nu = m_{a}c^{2} + E \Rightarrow h\,d\nu = dE \Rightarrow \frac{d\nu}{dE} = 1/h,
\end{equation}
and applying the change of variable in the distribution function, we finally obtain
\begin{equation}
\begin{aligned}
    f\left(\nu\right) & = 2\,\sqrt{\frac{\beta}{\pi}}\frac{h}{m_{a}\, v_{lab}}\, e^{-\beta v_{lab}^{2}\, -\, 2\beta E/m_{a}}\, \sinh\left(2\beta\, v_{lab}\,\sqrt{\frac{2E}{m_{a}}}\right)\\
    & = \frac{2}{\sqrt{\pi}}\,\sqrt{\frac{3}{2\langle v^2\rangle}}\,\frac{c^2}{\nu_{a}v_{lab}}\,e^{-\frac{3}{2\langle v^2\rangle}v_{lab}^2\, -\, \frac{3\left(h\nu\, -\, m_{a}c^2\right)}{h\nu_{a}\langle v^2\rangle/c^2}}\,\sinh\left(\frac{3}{\langle v^2\rangle}\,v_{lab}\,\sqrt{\frac{2\left(h\nu - m_{a}c^2\right)}{h\nu_{a}/c^2}}\right),
\end{aligned}
\end{equation}
where $\nu_{a}$ is the frequency of the motionless axion, and $\langle v^2\rangle$ $=$ $3\,k_{B}T/m_{a}$ is the root mean square velocity of the axions. Applying now that $\langle\beta^2\rangle$ $=$ $\langle v^2\rangle/c^2$ and $r$ $=$ $v_{lab}/\sqrt{\langle v^2\rangle}$, the result is the following:
\begin{equation}
    f\left(\nu\right) = \frac{2}{\sqrt{\pi}}\left(\sqrt{\frac{3}{2}}\frac{1}{r}\frac{1}{\nu_{a}\langle\beta^2\rangle}\right)\,\sinh\left(3r\sqrt{\frac{2\left(\nu - \nu_{a}\right)}{\nu_{a}\langle \beta^2\rangle}}\right)\,\mathrm{exp}\left(-\frac{3}{2}r^2\, -\, \frac{3\left(\nu - \nu_{a}\right)}{\nu_{a}\langle\beta^2\rangle}\right),
\end{equation}
which is the well-known expression for the frequency distribution of the axion in motion.

\section{Cross-correlation fundamentals}
\label{app:cross-correlation_appendix}

Cross-correlation is a mathematical operation that seeks for coincidences in two different signals, $s(t)$ and $g(t)$, and it is defined as:
\begin{equation}
    h\left(\tau\right) = s\left(t\right) \star g\left(t\right) \equiv \int_{-\infty}^{\infty}s\left(t\right)\cdot g\left(t + \tau\right)dt.
    \label{eq:cross_correlation}
\end{equation}
where $h(\tau)$ is the cross-correlation and $\tau$ is the variable of the shifts domain. One of its computational advantages is that, in Fourier domain, cross-correlation is just a product:
\begin{equation}
    \mathrm{FT}\left[h\left(\tau\right)\right] = \mathrm{FT}\left[s\left(t\right)\star g\left(t\right)\right] = \mathrm{FT}\left[s\left(t\right)\right]\cdot \mathrm{FT}\left[g\left(t\right)\right]^{*}.
\end{equation}
where FT makes reference to the Fourier Transform, defined as
\begin{equation}
    \mathrm{FT}\left[f\left(t\right)\right] = \int_{-\infty}^{+\infty}f\left(t\right)e^{-j\omega t}dt.
\end{equation}

The Power Spectral Density (PSD) of a given signal is directly related with its autocorrelation:
\begin{equation}
    \mathrm{PSD} = \frac{1}{T_{obs}}\left|x\left(\omega\right)\right|^2,
\end{equation}
where $T_{obs}$ is the observation time, and $x\left(\omega\right)$ is a signal in the frequency domain. In order to calculate the cross-correlation among signals, it will be computed by pairs. Thus, for three signals $S_1$, $S_2$, $S_3$ expressed in time domain, cross-correlation would have the following structure in Fourier domain: 
\begin{equation}
    \mathcal{H} = \mathrm{FT}\left[h\right] = \mathrm{FT}\left[S_{1}\right]\cdot \mathrm{FT}\left[S_{2}\right]^{*} + \mathrm{FT}\left[S_{2}\right]\cdot \mathrm{FT}\left[S_3\right]^{*} + \mathrm{FT}\left[S_{3}\right]\cdot \mathrm{FT}\left[S_1\right]^{*}.
\end{equation}

\begin{figure}[h]
    \centering
    \includegraphics[scale = 0.4]{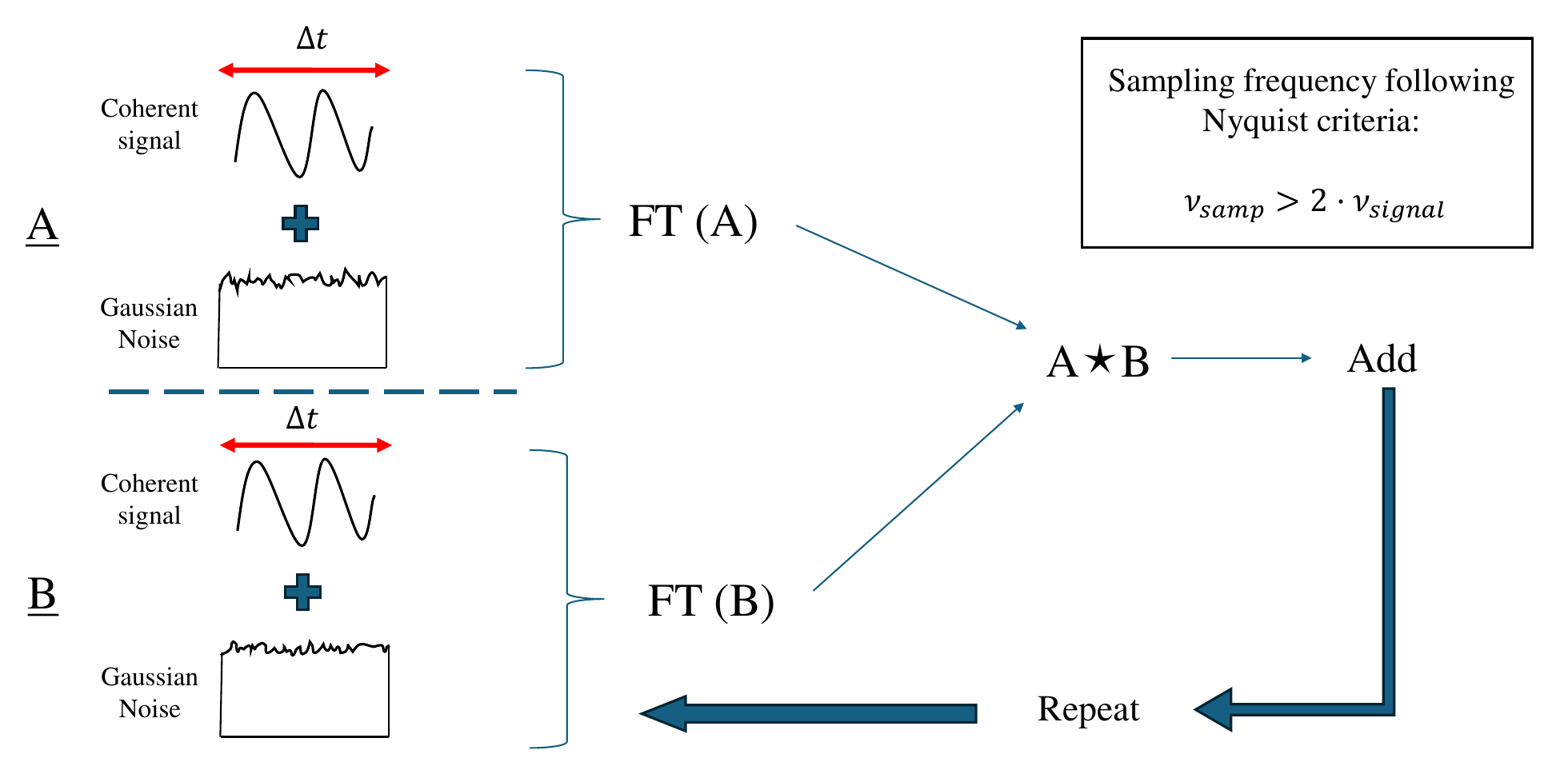}
    \caption{Sampling procedure of the signal from two hypothetical devices A and B. The signal coming from them is composed by noise plus a coherent signal. The full signal is portioned in small intervals of time duration $\Delta t$. After this, the Fourier transform is performed to both signals and next they are cross-correlated. The process is repeated, adding the last result to the previous one, until the exposure time of the experiment is reached.}
    \label{fig:sampling_process}
\end{figure}

In order to compute the cross-correlation, a sampling process of the signals must be developed. Let us have two devices named A and B, that represent two haloscopes. The signals coming from them are composed of noise plus the coherent signal of interest (the axion in our case), being sampled in small intervals of time length $\Delta t$. After this, the Fourier transform of the signals from both devices is numerically performed, and the result is then multiplied (i.e., cross-correlated in frequency domain). The result is saved and added to the next iteration of the process, until reaching the desired exposure time of the experiment. The explained procedure is schematically depicted in Figure \ref{fig:sampling_process}. The fundamental condition in order to reconstruct correctly the sampled signal is the Nyquist theorem \cite{Nyquist}.\\

During this work, we refer to time in terms of averages of short intervals (i.e., integration times). By averages, it refers to the number of signal segments with respect we perform the cross-correlation and the mean. This is more practical when trying to compare results with the bibliography, since the considered time for a given number of averages depends on the kind of data acquisition (DAQ) device available in the experimental setup (number of channels, acquisition velocity, sampling frequency, etc.).\\

\acknowledgments


This work was performed within the RADES collaboration. This work is part of the R\&D projects PID2022-137268NB-C53 and PID2022-137268NA-C55, funded by MICIU/AEI/10.13
039/501100011033 and by “ERDF/EU”. This work is also part of Quantum Technologies for Axion Dark Matter Search (DarkQuantum), funded by European Research Council (ERC-2023-SyG) (101118911). J. Reina-Valero has the support of ``Plan de Recuperación, Transformación y Resiliencia (PRTR) 2022 (ASFAE/2022/013)", funded by Conselleria d'Innovació, Universitats, Ciència i Societat Digital from Generalitat Valenciana (Spain), and NextGenerationEU from European Union. We thank to Prof. Mónica Argente Sanchis and Prof. José Joaquín Luján Soler for his useful comments and advices. Thanks also to Pablo Martín Luna for his valuable comments and discussions.



\bibliographystyle{JHEP}
\bibliography{biblio.bib}

\end{document}